\documentclass[authorversion,nonacm]{acmart}

\usepackage{amsthm}
\usepackage{amsmath}
\usepackage{graphicx}
\usepackage{multirow}
\usepackage{tabularx}
\usepackage{subfigure}

\AtBeginDocument{%
  }

\copyrightyear{2024}
\acmYear{2024}
\acmDOI{XXXXXXX.XXXXXXX}

\acmVolume{0}
\acmNumber{0}
\acmArticle{0}
\acmMonth{4}

\begin{document}

\title{Algorithmic Fairness: A Tolerance Perspective}

\author{Renqiang Luo}
\affiliation{%
	\department{School of Software}
	\institution{Dalian University of Technology}
	\city{Dalian}
	\postcode{116620}
	\country{China}}
\email{lrenqiang@outlook.com}

\author{Tao Tang}
\affiliation{%
	\department{Institute of Innovation, Science and Sustainability}
	\institution{Federation University Australia}
	\city{Ballarat}
	\postcode{VIC 3353}
	\country{Australia}}
\email{tao.tang@ieee.org}

\author{Feng Xia}
\authornote{Corresponding Author}
\affiliation{%
	\department{School of Computing Technologies}
	\institution{RMIT University}
	\city{Melbourne}
	\postcode{VIC 3000}
	\country{Australia}}
\email{f.xia@ieee.org}

\author{Jiaying Liu}
\affiliation{%
	\department{School of Economics and Management}
	\institution{Dalian University of Technology}
	\city{Dalian}
	\postcode{116024}
	\country{China}}
\email{jiayingliu@dlut.edu.cn}

\author{Chengpei Xu}
\affiliation{%
	\department{MIoT \& IPIN Lab, School of Minerals and Energy Resources Engineering}
	\institution{University of New South Wales}
	\city{Sydney}
	\postcode{NSW 2052}
	\country{Australia}}
\email{Chengpei.Xu@unsw.edu.au}

\author{Leo Yu Zhang}
\affiliation{%
	\department{School of Information and Communication Technology}
	\institution{Griffith University}
	\city{Gold Coast}
	\postcode{QLD 4222}
	\country{Australia}}
\email{leo.zhang@griffith.edu.au}

\author{Wei Xiang}
\affiliation{%
	\department{School of Computing, Engineering and Mathematical Sciences}
	\institution{La Trobe University}
	\city{Melbourne}
	\postcode{VIC 3086}
	\country{Australia}}
\email{w.xiang@latrobe.edu.au}

\author{Chengqi Zhang}
\affiliation{%
	\department{Australian Artificial Intelligence Institute, School of Computer Science, Faculty of Engineering and Information Technology}
	\institution{University of Technology Sydney}
	\city{Sydney}
	\postcode{NSW 2007}
	\country{Australia}}
\email{chengqi.zhang@uts.edu.au}

\renewcommand{\shortauthors}{Luo et al.}

\begin{abstract}
Recent advancements in machine learning and deep learning have brought algorithmic fairness into sharp focus, illuminating concerns over discriminatory decision making that negatively impacts certain individuals or groups. These concerns have manifested in legal, ethical, and societal challenges, including the erosion of trust in intelligent systems. In response, this survey delves into the existing literature on algorithmic fairness, specifically highlighting its multifaceted social consequences. We introduce a novel taxonomy based on 'tolerance', a term we define as the degree to which variations in fairness outcomes are acceptable, providing a structured approach to understanding the subtleties of fairness within algorithmic decisions. Our systematic review covers diverse industries, revealing critical insights into the balance between algorithmic decision making and social equity. By synthesizing these insights, we outline a series of emerging challenges and propose strategic directions for future research and policy making, with the goal of advancing the field towards more equitable algorithmic systems.


\end{abstract}




\maketitle

\section{Introduction}\label{sec:Intro}
With the increasing popularity of deep learning algorithms in automating social processes and promoting decision-making across various industry sectors, the prevalence of fairness issues in these algorithms has become increasingly apparent~\cite{saeid2022fairness}.
Unfairness refers to the tendency or prejudiced nature of decision-making exhibited by artificial intelligence (AI) systems, wherein they exhibit favoritism or discrimination towards individuals or groups based on specific sensitive features~\cite{ntoutsi2020bias}.
These sensitive features, such as gender, race, and region, are protected to prevent their misuse~\cite{garg2023handling}. Algorithmic fairness issues have significant implications for our daily lives, spanning healthcare~\cite{madhusoodanan2020racially}, industry~\cite{roselli2019managing}, education~\cite{guo2020graduate,xia2022summer}, medicine~\cite{roberts2021common}, social media~\cite{yu2022graph}, and policy-making~\cite{oyku2022fair}.
Instances of unfair algorithms have been observed in various domains, ranging from granting males low-interest loans~\cite{kallus2022assessing}, misleading perceptions about Black people on online platforms~\cite{kallus2022what}, skewing representations of women's online opinions~\cite{mehrotra2022fair}, or even causing delayed medical diagnoses for Black individuals~\cite{chen2023algorithmic}. Furthermore, there has been a growing concern within the research community regarding algorithmic fairness.

\textit{Fairness} is a concept rooted in value judgments and is a subjective evaluation, varying across cultures and societies~\cite{booth2021integrating}.
Fairness encompasses three key dimensions: distributive, interactional, and procedural fairness~\cite{cropanzano2001three}. Distributive fairness pertains to the perceived fairness in distributing consequential outcomes and allocating significant resources, such as employment opportunities. Interactional fairness focuses individuals' perceptions of the explanations, justifications, and justifications provided for organisational determinations. It also encompass individuals' appraisal of the interpersonal treatment they encounter during the decision-making process.
Procedural fairness centres on perceived equity in the various constituents of the decision-making process.

Current algorithmic fairness is not limited in legally protected sensitive attributes. It increasingly encompasses the equity of ethically significant judgments and the provision of personal rationales and justifications for algorithmic determinations~\cite{uddin2023spatio}. Recent literature reviews provide insights into algorithmic fairness by providing various definitions and exploring different fairness-aware methods. Dong et al.~\cite{dong2023fairness} provide a comprehensive overview of algorithmic fairness definitions in graph mining, encompassing group fairness, individual fairness, counterfactual fairness, degree-related fairness, and application-specific fairness. Wang et al.~\cite{wang2023a} present a definition of algorithmic fairness in recommender systems, including group fairness, individual fairness, and counterfactual fairness. Moreover, fairness in recommender systems involves calibrated fairness, envy-free fairness, Rawlsian maximin fairness, and maximin-shared fairness. Garg et al.~\cite{garg2023handling} illustrate definition of algorithmic fairness through the harm caused by fairness issues, encompassing sources of harm and targets of harm. Chen et al.~\cite{chen2023bias} review the sources of algorithmic fairness issues, highlighting bias in data, models, and results.
Bias refers to any systematic error in an algorithm that differentially affects the performance of different groups of users. Meanwhile, some reviews have focused on fairness-aware methods~\cite{pessach2022a, wan2023in}. Pessach and Shmueli~\cite{pessach2022a} provide an overview of algorithmic fairness, discussing its conceptual foundations and state-of-the-art approaches for enhancing fairness. They categorize fairness-aware mechanisms into pre-process, in-process, and post-process mechanisms.
Wan et al.~\cite{wan2023in} concentrate on in-processing methods, showcasing explicit mitigation and implicit mitigation.
However, these investigations overlook the ethical and personal perception aspects of algorithmic systems. 
The issue of fairness exhibits a more flexible tolerance within the framework of ethical norms and personal perceptions.

Distinct fairness-aware algorithms demonstrate varying degrees of tolerance.
The tolerance for fairness can be stratified at both the societal and individual levels as follows: zero tolerance by legal regulation, limited tolerance by ethical support, and flexible tolerance by personal perception. 
Legal regulations firmly reject discrimination and require the rigorous elimination of algorithmic bias that perpetuates social discrimination~\cite{kusner2020long}.
Fairness-aware methods aim to eradicate racial discrimination in judicial algorithms and gender discrimination in algorithms used by companies like Amazon for identification purposes~\cite{piano2020ethical}.
Furthermore, examining specific fairness should not be limited by legal regulation, as historical contexts of oppression, task nature, and contextual factors can influence model utilisation~\cite{mitchell2020algorithmic}.
Even in the absence of specific legal regulations, certain situations should limit the adverse outcomes of certain algorithms on ethical grounds~\cite{grossman2022media}.
For instance, within equitable competition, it is imperative to restrict algorithms that exhibit excessive favouritism towards a particular party to ensure a level playing field~\cite{grinberg2019fake}.
The acceptance of algorithmic explanations, rationales, and justifications can vary among individuals. 
Some individuals may rely heavily on recommendations from recommender systems, especially in online shopping or news, while others firmly believe that outcomes should not solely depend on these recommendations~\cite{abdollahpouri2019unfairness}.
Therefore, proposing a fairness taxonomy with varying tolerance levels is crucial for developing more precise strategies for fairness-aware methods.

\begin{figure*}
	\centering
	\includegraphics[width=0.9\textwidth]{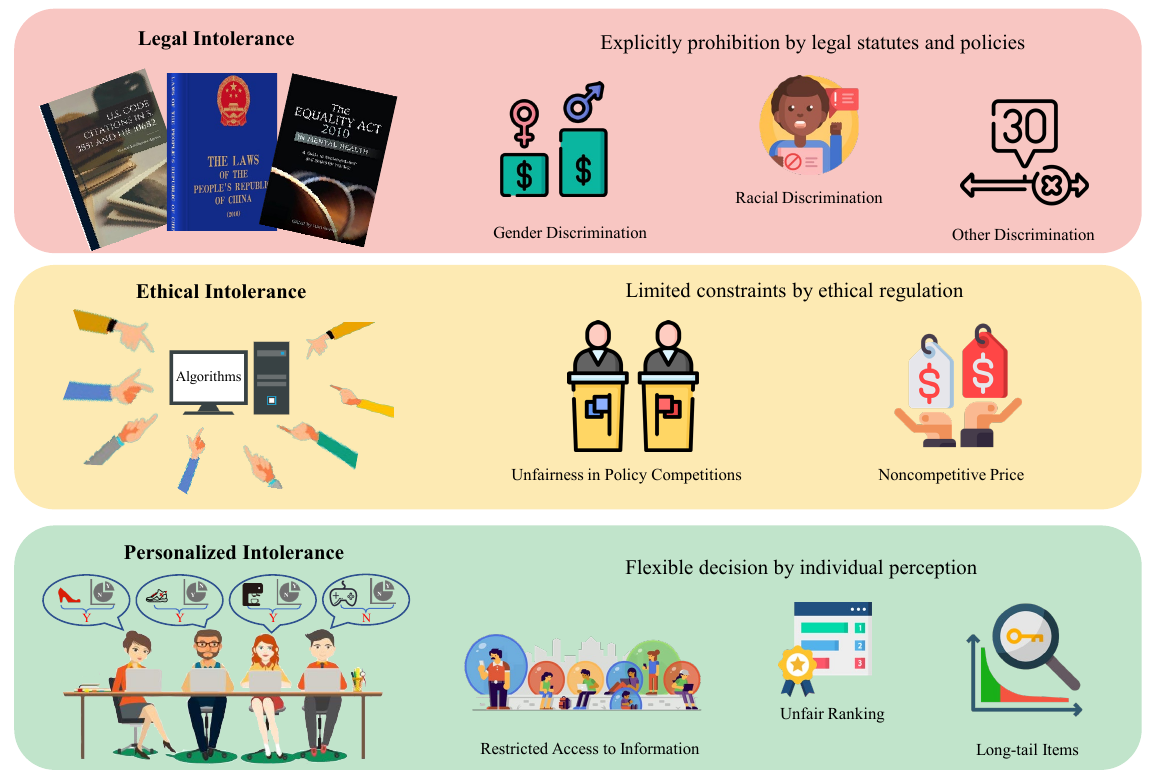}
	\vspace{-1ex}
  \caption{The tolerance and fairness issues.}
	\label{fig:TTAU}
  \vspace{-4ex}
\end{figure*}

This paper aims to investigate algorithmic fairness by exploring various tolerance levels: legal tolerance, ethical tolerance, and personal tolerance. 
First, legal tolerance pertains to discrimination that is explicitly prohibited by legal statutes and policies. 
Discrimination within the legal domain is typically categorized as direct discrimination, where differential treatment is intentionally based on an individual's membership in a protected class, or indirect discrimination, where seemingly neutral policies disproportionately impact members of protected classes~\cite{barocas2016big}.
The presence of either form of discrimination in algorithms necessitates a zero-tolerance approach.
Second, ethical tolerance focuses on limited constraints by ethical regulation.
For instance, during the 2016 U.S. presidential election, the influence of fake news on Twitter and the amplification of false information through recommendation algorithms exemplify algorithmic fairness issues~\cite{grinberg2019fake}.
Although clear legal regulations are currently lacking regarding issues such as algorithmic collusion and over-distribution, ethical resistance exists to these practices.
Third, personal tolerance is a flexible decision by individual perception.
While popularity bias in recommender systems has been established~\cite{zhang2021causal}, the acceptability of such bias varies among users with different levels of interest~\cite{abdollahpouri2021user}.
This type of fairness pertains to numerous individuals, each with unique requirements regarding the necessary extent of debiasing.
In Fig~\ref{fig:TTAU}, we summarize the examples of tolerance and fairness discussed in this paper. It is important to note that while we do not cover all existing fairness-aware methods, our focus is on a subset of them, examining different levels of tolerance.

The contributions of this work are summarised as follows:
\begin{itemize}
	\item We introduce a novel formulation of algorithmic fairness, viewing it through the lens of tolerance. By abstracting and synthesizing the concept of tolerance within algorithmic fairness, this review presents researchers with a fresh yet important perspective to mitigate prevalent fairness issues. This innovative taxonomy not only yields valuable insights but also catalyzes exploration of new avenues within this domain.
	\item Through systematic organization and analysis of diverse instances of algorithmic fairness, we provide a comprehensive overview that highlights the distinct degrees of tolerance involved. This systematic approach underscores the complexity of algorithmic fairness considerations and facilitates deeper understanding of its multifaceted nature.	
	\item We delve into challenging issues within this field and propose promising directions for future research. By identifying gaps and opportunities, we aim to stimulate further interest and innovation in addressing algorithmic fairness concerns.	
\end{itemize}

The remainder of the paper is organized as follows: We begin by discussing algorithmic fairness in relation to legal tolerance in Section~\ref{sec:LIG}. This is followed by an examination of fairness as regulated by ethical tolerance in Section~\ref{sec:EIB}. In Section~\ref{sec:PIB}, we explore fairness-aware methods grounded in personal tolerance. Challenges and future directions are then outlined in Section~\ref{sec:OCaI}, culminating with the conclusion of the paper in Section~\ref{sec:Con}.

\section{Legal Tolerance} \label{sec:LIG}
From a legal standpoint, algorithms are required to be free from any form of discrimination. 
Various countries have enacted legislation to prohibit discrimination, underscoring the critical need for equal treatment and the avoidance of bias based on protected characteristics. 
In the United States, for instance, discrimination on the basis of gender, age, disability, race, national origin, and religion is forbidden under Title 42, Chapter 21 of the U.S. Code. 
These laws cover a wide range of environments, including but not limited to education, employment, public access to businesses and buildings, federal services, and more.
The guidelines for human-AI interaction~\cite{amershi2019guidelines} underscore the importance of AI design in mitigating social discrimination. 
Real-world decision-making systems are increasingly becoming popular in processes such as rating, ranking and selecting/dis-selecting during online applications~\cite{zhang2021your}.
Without appropriate mitigation, pre-existing social stereotypes will be embedded in the algorithmic decision-making systems~\cite{silva2019algorithms}.
Algorithms should refrain from perpetuating the existing bias inherent in the data and actively work towards its elimination.

Despite these legal provisions, the representation of the less dominant gender continues to be inadequate and perpetuates stereotypical portrayals~\cite{costa2019analysis}. 
This issue is exemplified by the gender discrimination observed in Amazon's recruitment system, where historical bias in the training data has resulted in a significant bias towards male candidates and underrepresentation for female candidates~\cite{salimi2020database}.
Additionally, specific search algorithms exhibit racial discrimination by frequently triggering advertisements related to criminal records when encountering names associated with the Black community~\cite{chouldechova2020a}.
Despite the seemingly accurate predictive performance of using judicial identification as a proxy for determining guilt, considerable racial discrimination persists~\cite{zhang2022fairness}.
Table~\ref{tab:LIB} further lists more examples of the discrimination associated with natural language processing (NLP) techniques.

\begin{table}[H]
	\footnotesize
  \vspace{-4ex}
  \caption{Discrimination examples in algorithms.}
  \vspace{-4ex}
	\label{tab:LIB}
	\begin{tabular}{p{4.3cm}|p{4.3cm}|p{4.3cm}}
		\hline
		\textbf{Gender Discrimination} &
		\textbf{Race Discrimination} &
    \textbf{Other Discrimination} \\ 
    \hline
		Occupation words such as ``\textbf{professor}" and ``\textbf{nurse}" show discrepancy concerning the genders~\cite{poria2020beneath}. &
		The popular word embeddings is observed the negative association of \textbf{Black names}~\cite{caliskan2017semantics}.  &
    \textbf{Age-related} stereotypes are also embedding in the NLP community~\cite{Kathleen2022extracting}. \\ 
		\textbf{African American males} are far more likely to have their tweets detected as hate-speech than other users~\cite{kim2020intersectional}. &
		Researchers have noted a tendency for content created by \textbf{Black} users or utilizing \textbf{non-white} dialects to be more prone to mislabeling as toxic~\cite{garg2023handling}. &
    Some existing hate-speech detection methods trained some \textbf{sexual orientation words} as hate-speech~\cite{vaidya2020empirical}. \\ 
		Form English to Spanish, the stereotypical assignment of gender roles changes the meaning of translated sentences by changing the \textbf{gender} of the doctor~\cite{stanovsky2019evaluating}.&
		Most machine translation models are trained without the people who \textbf{do not speak English}~\cite{tan2020it}. &
    \textbf{Black man} and \textbf{Japanese man} have the most negative results in neural machine translation, and \textbf{Asian} and \textbf{Australian} have the most positive~\cite{wang2022measuring}. \\ 
    \hline
	\end{tabular}
  \vspace{-4ex}
\end{table}

\begin{figure}[H]
	\centering
	\subfigure[Reinforcing social stereotypes.]{
		\begin{minipage}[b]{0.45\textwidth}
			\label{RSS}
			\centering
			\includegraphics[width=1.0\textwidth]{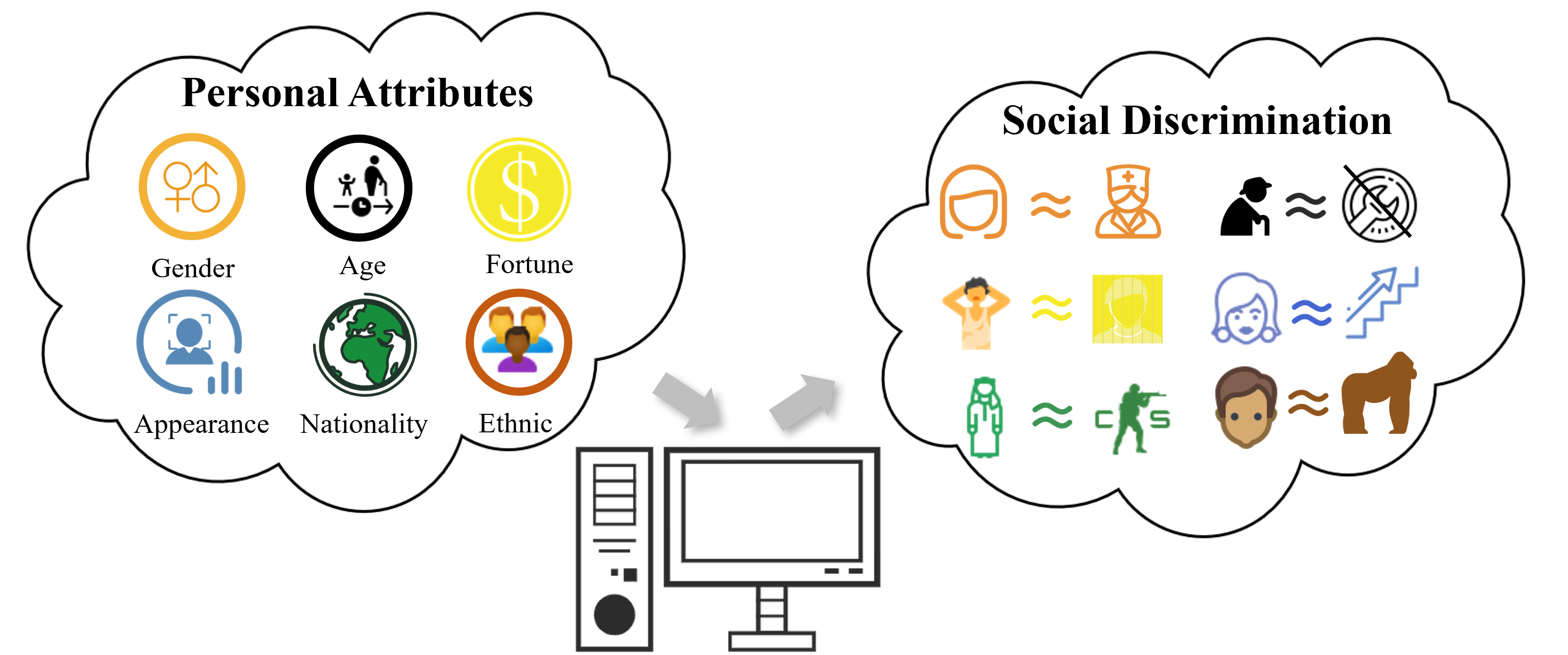}
		\end{minipage}
	}
	\subfigure [Causing inequitable distribution of resources.]{
		\begin{minipage}[b]{0.45\textwidth}
			\label{CIDR}
			\centering
			\includegraphics[width=1.0\textwidth]{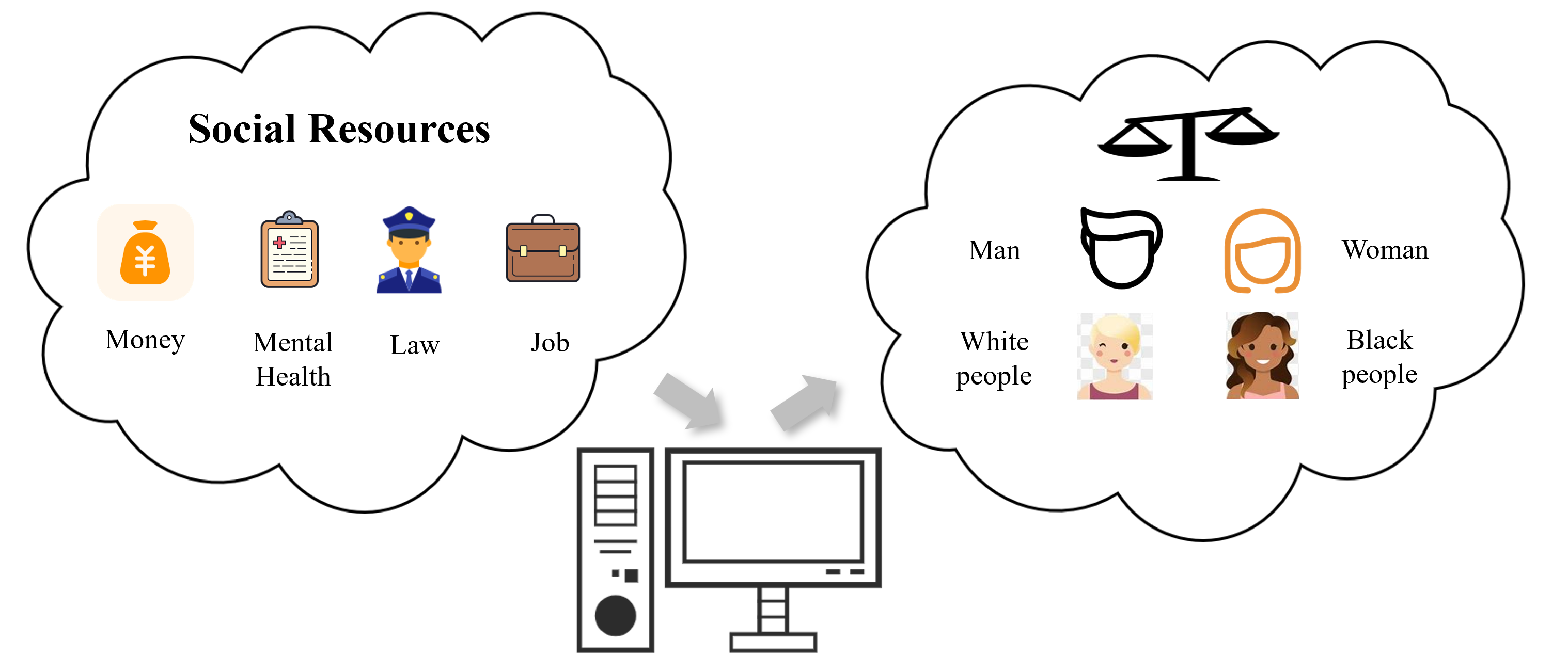}
		\end{minipage}
	}
	\caption{Algorithmic fairness issues exacerbating social discrimination.}
	\label{fig:UESD}
\end{figure}

Algorithms including healthcare algorithms, correctional offender management profiling for alternative sanctions (COMPAS) and various corporate hiring screening algorithms reinforce social stereotypes and unfairly allocate resources.
Gender and race are universally recognized as sensitive attributes protected by the laws of all nations, while attributes such as sexual orientation and religious beliefs are safeguarded by legislation in certain countries or regions.
With this background, we further classify discrimination into three types: gender discrimination, racial discrimination and other discrimination.
Furthermore, we illustrate the detrimental effects of fairness issues, including reinforcing (social) stereotypes and causing
inequitable distribution of resources, as depicted in Fig.~\ref{fig:UESD}.

\subsection{Gender Discrimination}
In recent years, several researches have revealed gender discrimination in algorithms.
Gender discrimination can occur when algorithms inadvertently perpetuate bias that is inherent in society, regardless of whether the algorithms themselves are intentionally discriminatory.~\cite{zhang2019faht}.
It is important to note that both gender discrimination originating from biased data and gender discrimination stemming from the algorithm's calculation process are deemed unacceptable.
The traditional way to limit gender discrimination, such as prejudice, relies on policy or law-enforcement.
In the context of current research, gender discrimination is commonly a binary classification problem.
Algorithmic gender discrimination is well handled when fairness is achieved in an exact binary classification setting~\cite{agarwal2018reductions}.

\noindent \textbf{Reinforcing Stereotypes:}
Word embedding, which captures relationships between words, has become an important part of NLP~\cite{xu2019lecture2note,hou2020network,xu2022semantic}.
People have stereotypical gender impressions of some words, and many NLP processes retain these stereotypes in their algorithms.
Hoyle et al.~\cite{hoyle2019unsupervised} study the language used to describe different genders. They point out significant differences between the descriptions of male and female nouns that are consistent with common gender stereotypes.
For example, positive adjectives used to describe women relate more to their bodies  than those used to describe men.
As one of the baseline methods in NLP, Hard-GloVe~\cite{bolukbasi2016man}, makes use of Global Vectors (GloVe)~\cite{pennington2014glove} to mitigate gender discrimination.
It consists of two steps.
The first step identifies and captures the subspace of the embedded bias.
The method uses analogies to quantify stereotypes $\mathbf{S}$ in embedding scores of all pairs of words $w_1$, $w_2$:
\begin{equation}\label{equ:EmbeddingScores}
    \begin{aligned}
        \mathbf{S}(w_1, w_2)= \begin{cases}\cos (\mathbf{s_1}-\mathbf{s_2}, w_1-w_2) & \text {if}\quad \|w_1-w_2\| \leq \delta \\ 0 & \text {otherwise}\end{cases}
    \end{aligned},
\end{equation}
where $(s_1, s_2)$ is a seed pair of words, such as $(s_1, s_2) = (\text{she},\text{he})$, $\delta$ is a threshold for similarity.
The second step ensures that the distance of gender-neutral words is 0 in the gender subspace, which is learned from gender specific words during embedding.
At the same time, this step equidistances the sets of words outside the gender subspace, forcing any neutral word to be equidistant from all words in each equal set.
For example, if $\{mother, father\}$ and $\{guy, gal\}$ are two equality sets, after the execution of second step, the word outside the gender subspace would be equidistant to $father$ and $mother$ and also equidistant to $guy$ and $gal$.

Based on Hard-GloVe, Zhao et al.~\cite{zhao2018learning} propose a Gender Neutral variant of GloVe, named GN-GloVe to isolate gender information without sacrificing the embedding quality by limiting gender information to sub-vectors.
The results present that the performance of GN-GloVe is better than that of Hard-GloVe in most word similarity tasks over the benchmark datasets. Furthermore, Based on GN-GloVe, Kaneko and Bollegala~\cite{kaneko2019gender} consider four types of information, including feminine, masculine, gender-neutral, and stereotypical, to show the link between gender and bias. They propose a bias-mitigating method named Gender-Preserving Glove (GP-GloVe).
The method obtains high accuracy in gender-definition (e.g.,``waiter-waitress") and decreases the gender-stereotype (e.g.,``doctor-nurse") in embedding.
Kumar et al.~\cite{kumar2020nurse} propose Repulsion, Attraction, and Neutralization-based debiasing Glove (RAN-GloVe), which is a fresh gender debiasing methodology that not only changes the spatial distribution of its neighbouring vectors but also eliminates gender discrimination contained in a word vector.
This method also maintains a minimum semantic offset to obtain the best performance in a downstream application task.
Though RAN-GloVe is inferior to GN-GloVe in terms of word embedding, it is the best method to mitigate the stereotype.
Wang et al.~\cite{wang2020double} propose the Double-hard GloVe for debiasing and study the advantages and disadvantages of the five algorithms discussed above.
It is concluded that the Double-Hard GloVe earns better F1-scores (\%) in both  OntoNotes and WinoBias datasets.
These word embedding methods have yielded substantial mitigation of gender stereotypes, thereby establishing a zero-tolerance approach towards their influence.

Reinforced gender stereotype also exists in machine translation (MT) systems. MT systems aim to automatically translate text or speech from one language to another.
In this process, MT models can significantly spread the gender discrimination.
For example, Hovy et al.~\cite{hovy2020can} use three commercially available MT systems (Google, Bing, and DeepL) in four languages (German, French, Italian, and Dutch) and reveal that there are significant differences in perceived demographics, with translations tending to masculine the author. To quantitatively study gender discrimination in MT, Stanovsky et al.~\cite{stanovsky2019evaluating} present a challenge set and an evaluation protocol named WinoMT.
This work is based on eight diverse target languages (Romance languages: Spanish, French, and Italian; Slavic languages: Russian and Ukrainian; Semitic languages: Hebrew and Arabic; Germanic languages: German), four tested popular commercial systems (Google Translate, Microsoft Translator, Amazon Translate, and SYSTRAN) and two recent state-of-the-art academic MT models (the best performance on English-to-French translation in the Conference on Machine Translation'14 test set, and the winner of the Conference on Machine Translation'18 for English-to-German translation).
They examine how these popular commercial systems and machine translation models correctly translated from English into $8$ different target languages with grammatical gender, and they discover that the tested models reinforce gender stereotypes.
Saunders et al.~\cite{saunders2020reducing} make a significant improvement by tuning a tiny, handcrafted fair gender dataset for three language pairs, and demonstrate better performance in comparison with all the systems evaluated in the research~\cite{stanovsky2019evaluating}. 
In their efforts to eliminate gender discrimination within the context of "black box" commercial systems, they are pursuing a comprehensive approach aimed at completely eradicating discrimination, striving for zero tolerance.

According to Kyriakou et al.~\cite{kyriakou2019fairness}, Computer Vision can be a source of gender discrimination by the way it tags images of people.
In a study conducted by Kyriakou et al.~\cite{kyriakou2019fairness}, the authors compare six taggers (Amazon Rekognition, Clarifai API, Google Cloud Vision, Imagga Auto-tagger API, Microsoft Vision, Watson Vision) on image search process and find that some platforms offer strange interpretations of images that may offensive and unfair to society. The bias is evident in the way the image taggers used gender related labels and also makes judgement about people's appearance. 
Since the tolerability of society bias is zero, the flaggers have removed offending tags from the database despite the so-called ``awkward workaround" employed by the flaggers.

A specific aspect within the realm of gender discrimination pertains to users' interaction with systems, particularly in the domain of information retrieval (IR).
Users vehemently oppose gender discrimination, prompting researchers to primarily focus on detecting user awareness of gender discrimination and elucidating the underlying reasons for its occurrence.
However, our understanding of how users perceive bias in search results, the extent to which these perceptions differ, and the extent to which they can be predicted based on user attributes remains limited~\cite{otterbacher2018investigating}.
Otterbacher et al.~\cite{otterbacher2018investigating} conduct an analysis to explore how users perceive bias, especially in terms of sexism, during image searching.
Through demographic analysis of the images with workers, they successfully detect the presence of gender discrimination.
The findings reveal that individuals with sexist beliefs exhibited a lower likelihood of detecting and reporting gender discrimination in image search results.
Additionally, Kopeinik et al.~\cite{kopeinik2023show} investigate whether and how users inadvertently replicate discrimination in their interactions with systems, thereby perpetuating the manifestation of bias.
Specifically, they explore the replication of stereotypical gender representations by users in their formulation of online search queries.
Their experimental findings confirm that users of search engines do replicate gender discrimination in their search queries.

\noindent \textbf{Inequitable Distribution of Resources:}
In addition to reinforced stereotyped algorithms, resource allocation can be biased when algorithms behave differently across different genders, recommending specific information related users' genders and unfairly allocating job opportunities~\cite{jiang2023fair}.

Some algorithms have better resolution in the male category than in the female category~\cite{liu2019data}.
For example, the coreference resolution algorithm, as a type of NLP algorithm, favours  masculine pronouns over feminine pronouns.
The reason can be attributed to the fact that feminine pronouns are significantly fewer than masculine pronouns in the text corpus.
This has an effect on development and training sets, ultimately resulting in the algorithm's incapacity to deal with female tasks.
To address the reinforcing stereotypes of gender discrimination in coreference resolution, Webster et al.~\cite{webster2018mind} propose a balanced corpus called Gendered Ambiguous Pronouns (GAP), which is a corpus of gender-balanced markers with a sample of $8,908$ ambiguous pronoun pairs to provide different coverage of the challenges posed by authentic texts.
This dataset mitigates the problem that existing datasets do not capture unambiguous pronouns with sufficient volume or diversity used in benchmark systems for practical applications.

Moreover, concerns regarding the footprint of gender discrimination in recommender systems have been raised~\cite{islam2021debiasing,barrio2019obtaining,stoica2020seeding}.
The recommendation systems serve as social networking tools to crowd-source gender-specific services to users, which can lead to inequality in representation and access to information~\cite{islam2021debiasing}.
Based on an analytical condition based on mathematical proofs, Stoica et al.~\cite{stoica2020seeding} describe a complex relationship between diversity and efficiency.
Based on this, they design and test an algorithm that uses network structure to optimize the dissemination of information while avoiding disparate impacts among participants of different genders.
Gender implications have been entirely eradicated.

Recommending specific information to users of different genders is an unfair allocation of resources because users may not receive similar information. For example, although male users have a higher preference for sports news, constantly recommending sports news in their news recommendations violates their right to obtain diverse news topics. To solve this problem, Barrio et al.~\cite{barrio2019obtaining} propose a random repair approach on balancing the ratio of access of information between genders.
They use the Wasserstein distance to blur the data, which is called total repair.
The random repair approach yields a tradeoff between minimal information loss and a certain amount of fairness. 
The objective is to entirely eliminate the impact of gender on the suggested outcomes.

Algorithms not only recommend specific information related to users' genders, but they also allocate job opportunities unfairly.
Resume search engines are information retrieval tools in which recruiters actively search for candidates based on keywords and filters.
Chen et al.~\cite{chen2018investigating} have examined whether the rankings exhibiting gender discrimination generated by search engines and discovered that professions like Software Engineer are unfair to women.
Ensuring fairness in these systems necessitates providing equitable success rates for protected groups.
In contrast to focusing solely on protected groups, Lahoti et al.~\cite{lahoti2019ifair} introduce a method that aims to achieve similar results for users with similar task-related attributes, regardless of potentially discriminatory attributes.
The method involves mapping user records' probabilities to a low-rank representation, which considers both individual fairness and the practicality of classifier and sorting in downstream applications.
By employing this method, users with similar job qualifications can receive similar recommendation rankings while disregarding gender.
Besides resume search engines, candidate slate for hiring recommendations will suffer from algorithmic bias, which breaks the uniform selection criteria for male and female candidates~\cite{peng2019what}.
Peng et al.~\cite{peng2019what} propose strategies to mitigate bias by disentangling the sources of bias, ensuring that gender has no influence on the outcomes generated by the system.

In conclusion, gender discrimination, being a prominent sensitive attribute, has garnered substantial research attention.
Despite extensive investigations, the prevalence of algorithms perpetuating gender discrimination persists, underscoring the need for continuous endeavors towards elimination.
Debiasing algorithms dedicated to combating sexism strive for complete eradication, encompassing the remediation of inherent stereotypes ingrained within the data and the prevention of emerging sexism propagated through algorithmic processes.
These efforts are driven by the commitment to uphold zero tolerance for gender discrimination.

\subsection{Racial Discrimination}
Racial discrimination is prevalent in the utilization of systems that rely on algorithmic decision-making. As a example, Rai~\cite{rai2019racial} highlight a concerning instance wherein black-skinned patients who received the same risk assessment from the algorithm were found in worse health compared to white-skinned patients.
The manifestation of racial discrimination within these systems not only perpetuates harmful stereotypes but also leads to the unfair distribution of resources. Such practices also run counter to legal principles that govern racial discrimination, further exacerbating the need for urgent attention and intervention.

\noindent \textbf{Reinforcing Stereotypes:}
We present a collection of articles that shed light on the phenomenon of racial discrimination reinforcing stereotypes.
Zou and Schiebinger~\cite{zou2018ai} sheds light on how specific word embeddings contribute to the perpetuation of social racial discrimination.
For instance, the analysis of Google Books data demonstrated that the word embeddings associated with attitudes towards Asian Americans in 1990 were characterized by terms such as ``inhibited" and ``sensitive".
Moreover, the authors emphasize that bias found in the data often stem from underlying institutional infrastructures and social power imbalances, which are deeply ingrained and not easily discernible.

Racial discrimination manifests within the domain of hate-speech detection, as the rapid expansion of online social language compels a heightened dependence on machine-driven screening methodologies.
Bagdaryan et al.~\cite{bagdasaryan2019differential} claim that differential privacy (DP), as a popular mechanism for training a model with privacy assurance, enhances the bias in NLP.
For example, the gender classification model trained by DP-SGD has much lower accuracy for black faces than white faces.
Their research focuses on unstructured data, in particular, which tackles the task of biased hate-speech detection.
Kim et al.~\cite{kim2020intersectional} discover systematic evidence of racial discrimination in datasets, including African American English (AAE) tweets of hate-speech and abusive language.
Hate-speech detection algorithms reinforce the bias of labelling AAE as abusive language.
AAE tweets are up to $3.7$ times more likely to be labelled as hate-speech, and AAE tweets from male users are up to $77\%$ more likely to be labelled as hate-speech than the others.
Sap et al.~\cite{sap2019the} also demonstrate racial discrimination in language detection in the Twitter dataset, and the bias could be traced back to the annotation process itself.
The toxicity score of AAE in the perspective application programming interface (API) is much higher than the tweets in non-AAE.
They employ dialect and race priming as methods to mitigate racial discrimination in annotation based on surface markers of African American English (AAE) ratings of toxicity. 
These approaches have led to the achievement of zero tolerance for racial discrimination.

However, it is important to note that bias is not limited to AAE alone. There are also biases present in the Hindi-English language, commonly referred to as Hinglish.
Hinglish is a portmanteau of Hindi and English, encompassing a blend of English and various South Asian languages spoken across the Indian subcontinent, often involving code-switching between these languages. Chopra et al.~\cite{chopra2020hindi} show a three-tier pipeline that employs profanity modelling, deep graph embeddings, and author profiling to retrieve Hindi-English language (Hinglish) hate-speech in Tweets.
They propose an expert-in-the-loop algorithm for debiasing the proposed model pipeline and investigate the generality and performance impact of bias elimination.
Although the automatic removal of hate speech reduces the growing pressure to employ such language on social media platforms, it also eliminates information from nodes with limited training, which can further silence already marginalized voices.
For example, Stoica et al.~\cite{stoica2018algorithmic} study the effect of racial dynamics on social recommendation through new Instagram data and discover that the platform hampers fair representation for groups like people of color.

In addition to deepening discrimination against AAE, some algorithms reinforce the stereotypes that black people are more prone to criminal behavior.
In particular, several articles focus on discrimination criminal defendant COMPAS, and the examples are outlined in Table~\ref{tab:COMPAS}.
Dressel and Farid~\cite{dressel2018accuracy} claim that though the COMPAS does not include race as an individual feature, other variables in the data may be correlated with race, resulting in racial discrimination. They discover that the COMPAS scores appeared to favour white defendants over black defendants because COMPAS underestimates recidivism rates for whites and overestimated recidivism rates for black defendants.
For debiasing in the COMPAS, Nabi and Shpitser~\cite{nabi2018fair} introduce modelling outcomes based on discrimination by ``sensitive characteristics".
They formalize racial discrimination as the presence of a certain path-specific effect (PSE) and minimize it.
Besides COMPAS, Mallari et al.~\cite{mallari2020look} demonstrate that the judgments of recidivism system also suffer from racial discrimination. The authors warn researchers to pay attention to gender and racial influences when using Amazon Mechanical Turk workers.
To solve the problem in the judgments of recidivism system, Metevier et al.~\cite{metevier2019offline} propose a versatile algorithm with multiple well-known and custom definitions of fairness.
The method is capable of producing unbiased strategies, and the performance is competitive with other offline and online context bandit algorithms.
Researchers have diligently endeavored to eliminate all racial influences embedded within algorithms that perpetuate and reinforce racial stereotypes, aiming to adhere to pertinent legal obligations.

\begin{table}[htbp]
	\small
  \centering
	\caption{The algorithmic fairness issues discovered in COMPAS.}
	\label{tab:COMPAS}
	\begin{tabular}{p{3.8cm}| p{8.2cm}}
		\hline
		\textbf{Discrimination} & \textbf{Examples of Algorithmic Bias in COMPAS}                                  \\ \hline
		\textbf{Gender Discrimination}          & The presented race of the defendant matter to \textbf{female} users~\cite{mallari2020look}.        \\ \hline
		\textbf{Racial Discrimination}          & The recidivism-risk is systematically overestimated for \textbf{black people}~\cite{piano2020ethical}. \\
		& \textbf{Black people} are almost four times more likely than \textbf{white people} to be arrested for drug offenses~\cite{dressel2018accuracy}.                                                      \\
		& Assuming the defendant is \textbf{Caucasian}, then the odds of recidivism for him would be 2.1 times more likely to reoffend if he/she is \textbf{African-American} (contrary to fact)~\cite{nabi2018fair}.\\
		\hline
	\end{tabular}
\end{table}

\noindent \textbf{Inequitable Distribution of Resources:}
Similar to the case of gender discrimination, algorithms also perform unfairly for different racial users.
For instance, the machine translation algorithm lacks multilingual diversity, which means the performance of algorithm in non-English is much worse than in English.
In particular, most of them are hard to run in the minority dialect language.
Jurgens et al.~\cite{jurgens2017incorporating} propose that debiasing in language identification (LID) should be the first step of machine translation.
They propose an updated dataset and a character-based sequence-to-sequence model to support dialects and multilingual language variants.
Tan et al.~\cite{tan2020it} expose the racial discrimination in popular machine translation models, such as bidirectional encoder representation from transformer (BERT) and transformer.
They tamper with the inflected forms of words to create plausible and semantically similar adversarial examples. 
They present that adversarial fine-tuning of individual epochs can significantly decrease racial discrimination without sacrificing clean data performance.

Unfair algorithmic performance also exists in the field of computer vision (CV).
Previous research have demonstrated that the presence of dark-skinned body in CV reflects implicit racial discrimination, leading to significant variations in algorithmic performance across different ethnic groups.
Specifically, Howard and Kennedy~\cite{howard2020robots} claim that autonomous vehicles exhibit poorer performance when trained with darker-skinned users, and commercial facial recognition application programs face the same problem.
They define a metric to examine the correlation between the performance of a deployed technology and a group's racial characteristics.
Wang et al.~\cite{wang2019racial} claim the Amazon Recognition Tool incorrectly matched photos of 28 congressmen to the criminal faces, especially among non-Caucasian people, with an error rate of 38\%. They compare four commercial application programming interfaces and four state-of-the-art algorithms.
Based on the findings, they propose a method using deep unsupervised domain adaption, where Caucasian is the source domain and other races are the target domains. 
Additionally, a deep information maximization adaptation network is introduced to reduce the racial discrimination.

Except for the performance of algorithms, some social resources face inequitable distribution because of racial discrimination in algorithms.
For example, resource allocation in health care is evaluated by comparable algorithms to achieve fairness and efficiency.
Nevertheless, Obermeyer et al.~\cite{obermeyer2019dissecting} uncover the evidence of racial discrimination in a widely utilized algorithm. 
Their study reveal that black patients who are assigned the same risk level by the algorithm are, in fact, sicker compared to their white counterparts.
They show an algorithmic remedy for this disparity in health care, which can increase the percentage of Black patients earning additional help from $17.7\%$ to $46.5\%$.
Unfortunately, recommender systems help spread racial discrimination due to their wide deployment~\cite{kallus2019the}.
Asplund et al.~\cite{asplund2020auditing} conduct a study using a controlled socks-puppet auditing technique to examine racial discrimination in the US Fair Housing Act. 
They focus on building online summary files associated with specific demographic summary files or cross-summary files.

Most home searches can now be conducted through property listing sites, home search engines, and targeted advertising, which can introduce subtle bias. For example, portraying an entry as a single racial entry situation can create a racial discrimination in subsequent searches and referrals.
Similarly, besides gender discrimination, hiring search engines are also prone to introduce racial discrimination.
Leung et al.~\cite{leung2020race} find that hiring search engines discriminate against black employees and favour Asian, female and attractive employees.
They also find that some users interface designs providing information about the candidate at the subgroup level, are also part of racial discrimination.
Hence, it is crucial to design the user interface of the platform in a manner that actively mitigates undesirable racial discrimination. 
This can be achieved by minimizing the disclosure of candidate-related information that may perpetuate biases based on race, thus achieving zero tolerance for racial discrimination.

\subsection{Other Discrimination}
Unlike race and gender, several sensitive attributes have obtained legal protection in the 21st century.
The UK government (Equality Act 2010: Guidance) has listed ``age, gender reassignment, being married or in a civil partnership, being pregnant or maternity leave, disability, disability, race including colour, nationality, ethnic or national origin, religion or belief, sex and sexual orientation" as the protected characteristics of discrimination.
It becomes crucial to enforce algorithmic mitigation strategies that effectively counteract the influence of particular sensitive attributes on the outcomes.
For instance, sensitive attributes such as age~\cite{zhang2018mitigating}, social hierarchy~\cite{chua2020you}, and political party affiliations~\cite{gordon2020studying} can introduce bias during the training process, leading to skewed results favoring a particular group.
Comments targeting different demographic groups can be dominated by the relationships that align with the majority group, leading to disparate performance~\cite{gutpa2023same}.
To mitigate polarization, it is necessary to remove these influence from these attributes.
For instance, the consideration of appearance in the context of job interviews necessitates careful attention and corrective measures~\cite{oneto2019taking}.

\noindent \textbf{Reinforcing Stereotypes:}
There is discrimination implicitly created by specific application sensitive attributes in downstream NLP applications.
Caliskan et al.~\cite{caliskan2017semantics} propose a purely statistical machine learning model named the word embedding association test (WEAT), which is based on replicating a spectrum of fairness. Their model discover the NLP applications biased by different nationality.
In formal terms, suppose that $\mathcal{W}_1$ and $\mathcal{W}_2$ are the sets of target words embedding of equal size, and $\mathcal{A}_1$ and $\mathcal{A}_2$ are the sets of attribute words, given a word embedding $\mathbf{e}_w$, the test statistic is:

\begin{equation} \label{equ:TestStatistic}
\begin{aligned}
d(\mathbf{e}_w, \mathcal{A}_1, \mathcal{A}_2) =&\text{mean}_{w_{1,i} \in \mathcal{W}_1} \cos (\mathbf{e_w}, \mathbf{w_{1,i}}) - \text{mean}_{w_{2,j} \in \mathcal{W}_2} \cos (\mathbf{e_w}, \mathbf{w_{2,j}}).
\end{aligned}
\end{equation}
For the query $\text{Q}=(\{\mathcal{W}_1,\mathcal{W}_2\},\{\mathcal{A}_1,\mathcal{A}_2\})$ and the embedding $\mathbf{e}$, the WEAT metric is defined as:

\begin{equation} \label{equ:WEAT}
\mathbf{WEAT}_{(\mathbf{e}, \text{Q})}=\sum_{\mathbf{e} \in \mathcal{W}_1} d(\mathbf{e}, \mathcal{A}_1, \mathcal{A}_2)-\sum_{\mathbf{e} \in \mathcal{W}_2} d(\mathbf{e}, \mathcal{A}_1, \mathcal{A}_2).
\end{equation}
where WEAT is a vector space based metric. Most approaches evaluate unintended demographic bias in the NLP community are based on this metric.
However, the reason for WEAT leading to bias in word embedding is unclear.
For interpretability, Sweeney et al.~\cite{sweeney2019transparent} provide a transparent framework to create a clear metric for measuring algorithmic bias in word embedding.
The metric, named relative negative sentiment bias, relies on the demographic identity terms relative to negative sentiment associated with various protected groups.
It allows researchers to obtain clear signals of algorithmic bias.

The detection of hate-speech, as a downstream NLP application, is also influenced by sexual orientation and religion.
Many existing hate-speech detection methods have been proposed to incorrectly associate non-hate-speech (e.g., lesbian, Muslim, gay) that have certain trigger-words as a potential hate-speech source~\cite{vaidya2020empirical}.
To address the issue, Vaidya et al.~\cite{vaidya2020empirical} propose a multi-task learning model consisting of an embedding layer, two long short-term memory network (LSTM) layers, an attention mechanism, and a custom loss function.
The custom loss function does not only aims to improve the overall ROC-AUC score on the test set but also places additional emphasis on mitigating unintended bias.
Dixon et al.~\cite{dixon2018measuring} manually curated a set of 50 terms after analyzing term usage patterns across the training dataset.
This selected set comprises terms with a similar nature but appear disproportionate occurrences in hateful comments.
For example, the word ``gay" appears in 3\% of hate-speech but only 0.5\% of comments overall.
While the detection of hate speech commonly employs demographic identity terms to measure bias, it often overlooks the protection of sensitive information for minority authors and readers.
In light of this, Badjatiya et al.~\cite{badjatiya2019stereotypical} introduce the concept of skewed occurrence measure to detect such bias-sensitive words and propose a new method that leverages knowledge-based generalizations.
This approach not only provides an abstraction that generalizes contents but also facilitates the extraction of information from the hate speech detection classifier.

\noindent \textbf{Inequitable Distribution of Resources:}
In algorithms for sentiment prediction, the algorithm can produce bias in prediction performance due to attributes such as age, nationality and disability.
For example, the original review is ``there are a lot of actors trying too hard to be funny", but the biased reviews including ``there are a lot of actresses trying too hard to be funny" and ``there are a lot of Chinese actors trying too hard to be funny" are appeared~\cite{ma2020metamorphic}.
The sentiment prediction result of the first review is unhappy, but the predictions of the other two reviews are happy.
When some unemotional attributes (such as gender and nationality) change or are explicitly specified in the sample, the results of affective predictions will change, leading to potential unconscious bias.
Ma et al.~\cite{ma2020metamorphic} flag thousands of discriminatory inputs by a well-established software testing scheme.
The testing scheme enhances the fairness guarantee at a modest cost.
Similarly, Diaz et al.~\cite{diaz2018addressing} discover a significant age bias across $15$ sentiment analysis models.
For example, an adjective with the word ``young" is $66\%$ more likely to receive a positive score than the same sentence with the adjective ``old".
They propose a method with a customized classifier to decrease irrelevant attribute bias by an order of magnitude.
Besides nationality and age, disability is another irrelevant attribute that influences fair competition.
More than one billion people (about $15\%$ of the world's population) are disabled, and some sentiment predictions treat disability as the subject of strong negative social prejudice.

Facial expression recognition is another field suffering discrimination from age.
The demographic attributes such as age is inequitable distribution of the population in most face image datasets.
Xu et al.~\cite{xu2020investigating} conduct a systematic study on bias and fairness in facial expression recognition on two well-known datasets, namely RAF-DB and CelebA, through a comparison of three different methods including the baseline, the attribute-aware method, and the disentangled method.
They discover that the attribute-aware approach and the disentangled approach,  both equipped with data augmentation, outperform the baseline approach in terms of accuracy and fairness. The disentangled approach performs best in reducing demographic bias.
Amini et al.~\cite{amini2019uncovering} propose a tunable algorithm for reducing algorithmic bias within the training data.
The algorithm adjusts the respective sampling probabilities of individual data points for training.
By learning irrelevant attributes in a completely unsupervised manner, the algorithm decreases the bias of irrelevant attributes in the training set without manually labelling.

Several researchers have focused their efforts on striking a balance between algorithmic accuracy and the presence of algorithmic bias by avoiding sensitive attributes in specific application contexts.
It is crucial to comply with legal considerations and minimize the reliance on such attributes to mitigate the introduction of avoidable bias.
Oneto et al.~\cite{oneto2019taking} propose a three-pronged method by using Multitask Learning: increasing accuracy in each sensitive group, enhancing measures of fairness, and protecting sensitive information.
This method treats people belonging to different sensitive groups but with similar non-sensitive attributes equally.
Krasanakis et al.~\cite{krasanakis2018adaptive} propose a theoretical basis for the model to iteratively adapt the training sample weighting process, which mitigates different types of bias and achieves better fairness.
The disparate mistreatment elimination constraint is to equalize false positive rates and equal false negative rates among sensitive and non-sensitive groups:

\begin{equation} \label{equ:FPRandFNR}
\begin{aligned}
    &\text{P}(\hat{l}_o \neq l_o | l_o = 1, o \in \mathcal{S})=\text{P}(\hat{l}_o \neq l_o | l_o = 1, o \notin \mathcal{S}), \\
    &\text{P}(\hat{l}_o \neq l_o | l_o = 1, o \in \mathcal{S})=\text{P}(\hat{l}_o \neq l_o | l_o = 1, o \notin \mathcal{S}),
\end{aligned}
\end{equation}
where the $\hat{l}_o \in \{0, 1\}$ reflects the label estimation for outcome $o$, $l_o \in \{0, 1\}$ reflects the label for outcome $o$, and $\mathcal{S}$ is the sensitive attribute set.
These methodologies not only uphold high algorithmic accuracy but also safeguard the confidentiality and integrity of sensitive attributes associated with users in the specific application domain.

In summary, algorithmic bias can produce fairness issues in two main ways: the first reinforces some of the original bias, and the second produces an unfair distribution for different social groups in the process.
It is worth noting that most articles have discussed binary dimensions, such as old and young, hate-speech and non-hate-speech, even social class defined by a baseline.
Current algorithms cannot handle multivariable bias simultaneously.

\section{Ethical Tolerance} \label{sec:EIB}
Although several organizations have published ethical checklists outlining principles for algorithmic decision-making systems, concerns persist regarding their potential misuse~\cite{krafft2021action}.
Researchers are anxious that algorithmic decision-making systems can be misused by malevolent actors, thus eluding accountability or inadvertently disseminating bias and harming the competitive interests~\cite{jobin2019global}.
Particularly concerning is the need for more censorship in the real-world deployment of these systems.
Even if the training samples are representative and accurate, they may still record objectionable aspects of the world that, if encoded in a policy, run counter to the decision-makers goals~\cite{mitchell2020algorithmic}.
For example, social media is reported to be ideal for spreading unconfirmed or biased information~\cite{baeza2020bias}.
Confirmation bias and filter bubbles driven by algorithmic bias further amplify the spread of fake news, exacerbating the manipulation of public opinion by profit-driven individuals and even influencing political elections~\cite{pierri2020topology}.
Except for the 2016 U.S. election, which most researchers have already paid high attention to, the 2018 Italian political election is also influenced by algorithm bias~\cite{becatti2019extracting}.
The unfair algorithm also leads to the amplification of rumours about COVID-19 on social media~\cite{yu2021data}.

Meanwhile, the competition in the market system is also influenced by algorithms.
Unhealthy competition is collusion, where businesses agree not to compete with each other so they do not need to charge lower prices in an effort to attract customers~\cite{mancini2022self}.
The problem is called non-competitive price.
Calvano et al.~\cite{calvano2020protecting} claim that such collusion occurs when price-setting algorithms learn to adopt collusive pricing rules without human intervention, oversight or even knowledge.
Ethics-aware methods mitigates adverse consequences through constraining algorithmic outcomes.
Researchers, as well as users of the algorithms, also aim to achieve limited constraints through ethical regulation.

\subsection{Unfairness in Policy Competitions}
Social media is a peculiar route for news spreading, information exchanging, and fact checking~\cite{caldarelli2020the}.
The rapid widespread dissemination of activism has become an increasingly severe problem with the development of social media~\cite{sanford2022veganuary}.
Because of the explosion of information, platforms tend to use automated decision-making systems to help recommend and filter information~\cite{jiang2023trade}.
However, automation increases the spread of inflammatory content on social media platforms~\cite{jahanbakhsh2022leveraging}.
These contents ranges from the partisanship and misinformation to influence the government election~\cite{artime2020effectiveness}.

The task of reducing partisanship opinions and mitigating hate speech is multifaceted and requires a comprehensive approach. However, it is important to acknowledge that certain algorithms may inadvertently contribute to the spread of hate speech while attempting to decrease partisanship opinions, whereas others may inadvertently amplify partisanship opinions while aiming to mitigate hate speech.
Munn~\cite{munn2020angry} claim that Facebook and YouTube privilege inflammatory content through a stimulus-response loop that promotes angry utterances.
The algorithms on both platforms are biased towards hate-speech.
To address the spread of hate-speech, social media platforms use a mix of manual and algorithmic processes to implement content auditing~\cite{klonick2020content}.
However, the content moderation algorithm, which aims to remove hate-speech from social media, has attracted the attention of lawmakers due to allegations of partisanship~\cite{fan2020digital}.
Despite the absence of existing legal regulations, scholars have undertaken endeavors to devise methodologies aiming at mitigating the adverse consequences associated with these algorithms, motivated by ethical considerations.
Jiang et al.~\cite{jiang2020reasoning} propose two formal criteria (i.e., independence and separation) to measure political bias in the hate-speech detection algorithms on YouTube.
The independence is based on the demographic parity, and the separation is based on equalized odds.
For debiasing the political bias in hate-speech mitigation algorithms, researchers call for transparency in content moderation practices and recommend that platforms preserve and protect the moderated content.
The fairness of the algorithm is constrained by certain ethical limitations.

The identification of factors that warrant ethical restraint is essential for addressing prejudice within the framework of ethical guidelines.
A successful ideology-detection is the first step to mitigate algorithmic political bias~\cite{coates2018unified}.
Significant segments of the population share opinions and news related to the politics or causes they support, thus providing solid hints of their political preferences and ideologies~\cite{preoctiuc2017beyond}.
Preotiuc-Pietro et al.~\cite{preoctiuc2017beyond} focus on politically moderate and neutral citizens.
They aim to build a fine-grained model that mines the invisible political ideology of users.
Concerning ideology-detection, researchers commonly conceptualize ideology as a single, left-right binary problem.
Their work contributes to studies of political bias in the sense how ideology is revealed across users with various levels of engagement.
Within certain policy frameworks, politicians can express their stance on issues and selectively use certain political slogans on Twitter to display their underlying political ideology.
Johnson and Goldwasser~\cite{johnson2018classification} have paid attention to the policy frames correlated with political ideologies.
They proposed probabilistic soft logic, a graphical probabilistic modelling framework, which discover the political ideologies from politicians' policy frames.
Similarly, Xiao et al.~\cite{xiao2020timme} propose Twitter ideology-detection via multi-task multi-relational embedding (TIMME), which is efficient on a sparsely-labelled heterogeneous real-world dataset.
Experimental results show that TIMME generally outperforms the current graph convolutional network (GCN) and R-GCN for ideology-detection.
In theory, TIMME can be extended to other datasets and tasks.

Meanwhile, some researchers mine political bias without ideology-detection.
Hu et al.~\cite{hu2019auditing} build a dictionary of partisan cues to measure partisanship and a set of left-learning and right-learning search queries.
They collect a large dataset of search engine results page (SERPs) on Google Search by running their search queries and their autocomplete suggestions.
They find that Google Search code snippets typically amplify partisanship.
Ye and Skiena~\cite{ye2019mediarank} design an automated rank system, called MediaRank, to analyze the partisanship influence of media sources in four English-speaking countries.
MediaRank collects and analyses $1,000,000$ news web pages and $2,000,000$ related tweets every day to rank $50,000$ online news sources worldwide.
They discover that, in four English-speaking countries (the United States, the United Kingdom, Canada, and Australia), the highest-ranked news sources all disproportionately support left-wing parties.

In conclusion, the presence of politically biased content on platforms hinders the formation of users' opinions. The algorithms play a significant role in amplifying dissemination.
While political competition is a valid aspect of democratic processes, it is undesirable for political parties to exploit algorithms to disrupt the equilibrium of competition.
The challenge lies in detecting and addressing implicit forms of algorithmic bias that perpetuate political bias.
Therefore, the central concern revolves around devising effective regulatory mechanisms based on ethical standards to uncover and mitigate potential political bias, thereby addressing fairness issues.

\subsection{Noncompetitive Price}
\begin{figure}
	\centering
	\includegraphics[width=0.5\textwidth]{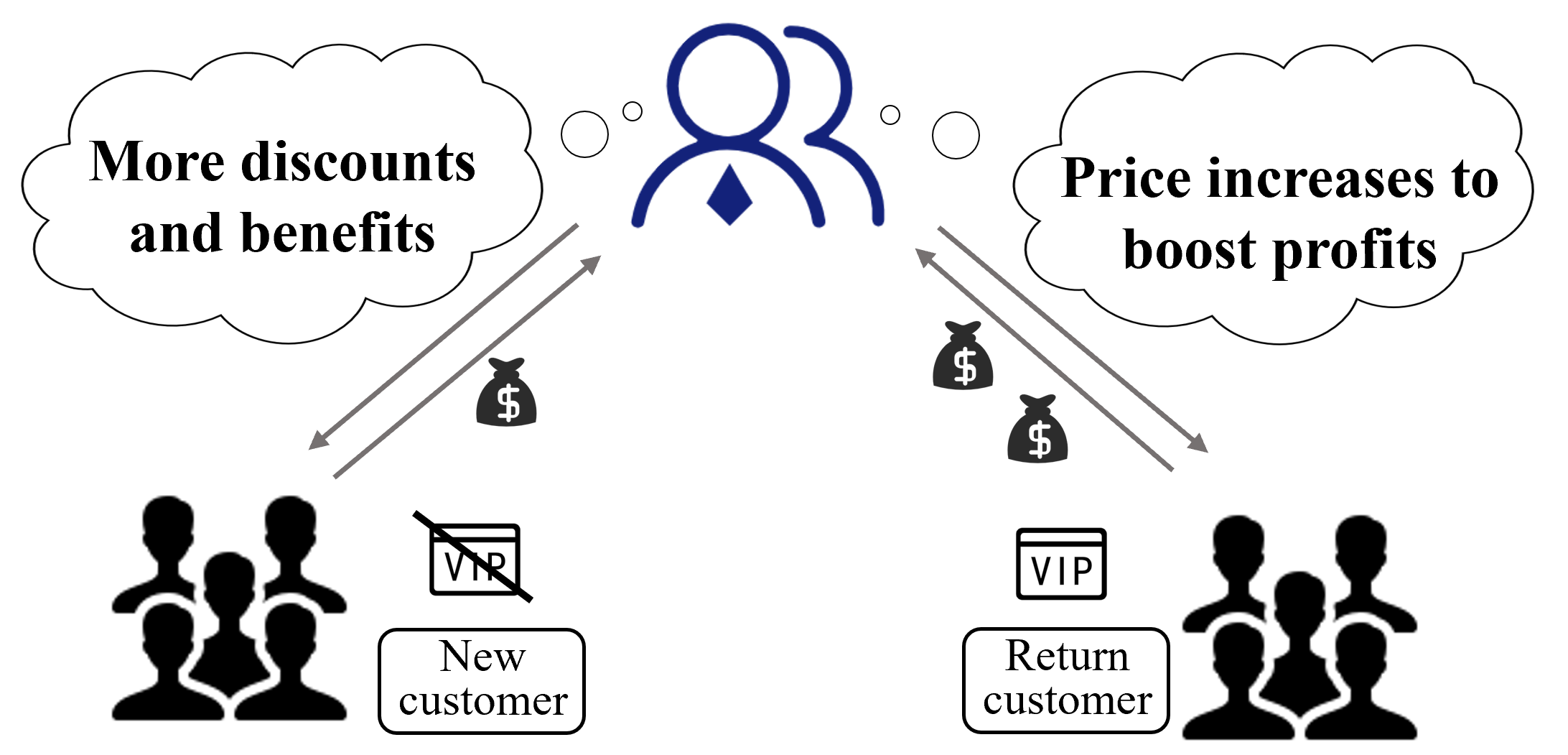}
	\caption{Noncompetitive price.}
	\label{fig:NP}
\end{figure}

Price-setting algorithms' bias, also called noncompetitive prices, has received extensive attention from the media, industry, and regulatory agencies ~\cite{cohen2019price}.
Co-operations use algorithms for marketing budget allocation~\cite{zhao2019unified} and price-setting~\cite{feng2019smart}.
Designed solely for making profits, algorithms may inevitably deviate from the public interest.
Information asymmetries, bargaining power, and externalities are prevalent in markets.
The performance of noncompetitive price is shown in Fig~\ref{fig:NP}.
Very important person (VIP) members or returning members spend more money than new customers and non-VIP members on the same items. In addition, algorithms mark different prices for the same consumed products on different terminals.
For example, users in communities with higher proportions of non-whites tend to pay more for ride-hailing services~\cite{schor2021the}.

Miklos-Thal and Tucker~\cite{miklos2019collusion} build a game-theoretic model to examine how price-decision algorithms affects algorithmic collusion.
While price-decision algorithms allow colluding firms to better adjust prices to demand conditions, they also encourage each firm to deviate from the lower price during periods of high forecast demand.
Using price-decision algorithms for pricing by businesses has implications for consumer welfare.

The noncompetitive price has recently become a new frontier in competition policy and anti-trust regulation.
Calvano et al.~\cite{calvano2021algorithmic} claim that companies could use social network information to inform their pricing decisions.
They show how much value companies can be added if the companies make information public.
Meanwhile, the public pricing decisions analyze whether the consumers get a higher price than their peers.
Some price-setting algorithms based on Q-learning, are originally proposed to tackle Markov decision processes.
Their study shows that algorithmic collusion is not only an artificial behavior among collaborators, but may also be the result of automatic evolution in algorithmic computing.
Their investigation shows that the Q-learning algorithm converges on cooperative behaviour in the simple prisoner's dilemma.
The algorithm is confirmed in complex and realistic environments.

Calvano et al.~\cite{calvano2019algorithmic} also find that the Q-learning algorithms are constantly learning to charge hyper-competitive prices without talking to each other. 
The researchers conducted numerical simulations within a synthetic environment, where multiple cycles of interaction between the same Q-learning algorithms took place. 
The findings reveal that algorithms can collude even in the absence of explicit instructions or direct communication with each other, under conditions of imperfect monitoring.

Regulators need tools to help detect such collusion.
In reality, researchers only have access to the agent's strategies and external shock value, while the agent's utility function is unknown.
Hespanhol and Aswani~\cite{hespanhol2020hypothesis} propose a hypothesis testing framework to decide whether these kinds of agents are behaving competitively.
In their setting, agents are utility-maximizing and compete over the prices of items.
They use the formulation without knowing the agent utility to estimate the unknown private information vectors and approximate equilibrium residuals.
Then, they confirm their method with computational experiments on the Bertrand competition game.
In particular, their settings are flexible because the regulator does not require the agent's utility function.

Algorithm developers bear the responsibility of ensuring that their algorithms maintain an objective stance, supporting both sides of the competition to uphold fairness and prevent unjust competition.
While this ethical principle is widely acknowledged, the practical implementation of algorithms often veers away from these fairness ideals, concealing inherent bias.
Current research efforts focus on identifying and addressing these inequalities and achieving equity through the development and implementation of relevant policies based on ethical regulation.

\section{Personal Tolerance} \label{sec:PIB}
As algorithms continue to advance in their ability to predict individual behavior, there is a growing trend of people relying on algorithmic recommendations in various aspects of their lives~\cite{wang2019urban}.
People have the right of choosing information recommended by systems, and they also demand diversity in this information~\cite{li2020community}.
Individuals' popularity tendencies show different trends. 
The algorithms often show a clear preference for popular items~\cite{smith2023the}.
Notably, this bias can disproportionately affect highly engaged, discerning, and challenging-to-predict users, leading to an fairness issues concern in their recommendation~\cite{yalcin2022evaluating}.
For example, a Big Bang fan may feel unhappy if the system uninterruptedly recommends the series~\cite{rhee2022countering}.

It is worth noting that unfair ranking can also influence the recommendations received by merchants~\cite{zhao2022investigating}.
Most ranking algorithms are trained with only displayed items, most of which are popular, It exacerbates the neglect of non-displayed items~\cite{zhang2022incorporating}.
This phenomenon decreases the diversity of items for consumers and the business of non-displayed merchants~\cite{chen2022co}.
More extreme projects that are not shown are called long-tail projects~\cite{zhao2023popularity}.
It's crucial to acknowledge that the tolerance for algorithmic fairness is highly individualized.

\subsection{Restricted Access to Information}
Recommender algorithms also result in users' restricted access to information, creating the ``echo chamber"~\cite{michela2016echo} and the "filter bubble"~\cite{tien2014exploring}.
These two terms have a similar meaning in the sense that recommender algorithms enhance the user's interests through repeating the exposure to similar contents. But they prevent users from engaging with ideas different from their own, and limit access to information.
The ``echo chamber" mainly emphasizes the generated population, while the ``filter bubble" is concerned with the limiting of information.
One reason for producing this negative impact is the ``confirmation bias".
Due to bias, people prefer to read content that is the same as what they believe and ignore content that is different.
Brugnoli et al.~\cite{brugnoli2019recursive} quantitatively investigate the impact of two mechanisms (challenge avoidance and reinforcement seeking) of the ``confirmation bias".
They confirm the relevance between ``confirmation bias" and ``echo chamber".
The ``confirmation bias" trigger the emergence of two distinct and polarized groups of users (i.e., ``echo chamber"), whose surrounding friends also tend to share similar systems of beliefs.
Similarly, the ``Filter bubble" links user polarization to algorithmic filtering, leading users to assemble very similar briefs ~\cite{pariser2011filter}.
The target users only get recommendations for the most familiar entries and lose links to many other available entries.
The examples are shown in Table~\ref{tab:RestrictedAcesstoInformation}.

\begin{table}[]
	\small
  \centering
	\caption{The examples of restricted access to information.}
	\label{tab:RestrictedAcesstoInformation}
	\begin{tabular}{p{2.5cm}| p{9.5cm}}
		\hline
		\textbf{Type of bias} & \textbf{Example}                                                                                               \\ \hline
		\textbf{Echo Chamber}   & Personalized e-commerce has a tendency to \textbf{``echo chamber"} on user click behavior in Alibaba Taobao~\cite{ge2020understanding}.                \\
    \\
		\textbf{}               & This can lead to social fragmentation and fragmentation online \textbf{``echo chambers"}, where the same ideas keep reverberating over and over again, potentially exacerbating the effects of \textbf{``confirmation bias"} as users may redistribute the same information to which they are selectively exposed~\cite{sikder2020a}.  \\
    \\
		\textbf{}               & The \textbf{``echo chamber"} for the websites publishing factually dubious content are often described as fake news in the 2016 U.S election~\cite{guess2020exposure}.
		\\ \hline
		\textbf{Filter Bubber} &
		Because of the accounts users follow and the friends they have on social media, they receive news with a \textbf{specific political bias}~\cite{bentley2019understanding}. \\
    \\
		\textbf{}               & The search results are \textbf{limited to homogeneous information} on YouTube~\cite{hussein2020measuring}, Twitter~\cite{c2020there}, and Google Search~\cite{robertson2018auditing}.
		\\ \hline
	\end{tabular}
\end{table}

\subsubsection{Echo Chamber}

The premise of eliminating the ``echo chamber" is to discover it.
Ge et al.~\cite {ge2020understanding} discover the ``echo chamber" in two steps: first, they explore whether the user's interest was enhanced; second, they check whether the reinforcement came from exposure to similar content.
They cluster users into multiple groups based on preferences in the first step, and use cluster validity metrics to measure changes in user interests.
They measure the diversity of content in the recommendation lists in the second step, which studies whether recommender systems narrow down the range of content offered to users.
Their assessments are enhanced by robust indicators, including cluster validity and statistical significance.
Sikder et al.~\cite{sikder2020a} introduce a social learning model for discovering the ``echo chamber".
In the model, most participants update their beliefs based on new information, while a minority of participants reject information inconsistent with their existing beliefs.
This simple mechanism generates the ``echo chamber".
The results explain the trade-off between the ``echo chamber" and social connectivity.
Society is more likely to maintain a long-term narrative representing all available information. 
However, such a society may become more polarized.

Unlike legal tolerance and moral tolerance, the mitigation of personal tolerance is usually approached with a more flexible perspective.
Rather than seeking to completely eliminate the effects of specific attributes, researchers have proposed complementary measures to mitigate the detrimental consequences of this bias.
Rather than seeking to completely eliminate the influence of a specific attribute, researchers propose supplementary measures to mitigate the adverse consequences stemming from such bias.
One helpful method to solve the ``echo chamber" is generating more diverse and relevant recommendations~\cite{wan2019to}.
Wu et al.~\cite{wu2019pd} propose a new recommendation model based on Personalized Diversity promoting generative adversarial network (PD-GAN), to increase the diversity of recommendations.
It is an adversarial generator based on the general co-occurrence of various items and the users' personal preferences for items.
The experimental results show that PD-GAN has superior performance in generating relevance and diversity recommendations than the baseline methods.
Similarly, Sun et al.~\cite{sun2019debiasing} propose a novel blind-spot-aware matrix factorization (MF) algorithm to generate more diverse and relevant recommendations.
MF algorithm mitigates the phenomenon that users only see a narrow subset of the broad range of available recommendations.
The ``echo chamber" can lead recommender systems to degenerate into polarized.
There exist methods aiming at slowing down system degeneracy.
Jiang et al.~\cite{jiang2019degenerate} provide a theoretical analysis to examine the effects of user dynamics and the recommender system.
They use the dynamical model framework to model users' interests and treat the extremes of interests as the degeneracy points of the system.
Following the proposed framework, they formally define what system degradation is and examine how to mitigate it.

\subsubsection{Filter Bubble}

Some researchers study to confirm or assess the impact of the ``filter bubble".
Chitra et al.~\cite{chitra2020analyzing} propose a mathematical framework to assess the impact of the ``filter bubble" based on the Friedkin-Johnsen opinion dynamics model.
They confirm the ``filter bubble" and explain why social networks are vulnerable to outside actors.
Bently et al.~\cite{bentley2019understanding} observe the ``filter bubble" in the new habits of 174 American participants.
Nearly half ($48\%$) of the participants have received left-biased sources overwhelmingly, while only $5\%$ have received right-biased sources.
People are not exposed to different viewpoints, though the population overall is learning farther left.
This phenomenon confirms the ``filter bubble", where people don't understand the perspectives of other groups of people.
Many users worry that the results of their searches may not be representative.
Stoica and Chaintreau~\cite{stoica2019hegemony} describe an organic growth model to investigate whether social media and recommender systems pay attention to different points of view.
The biased preferential attachment model inspires the organic growth model, including minority-majority partition, item creation, reposting, and homophily.
Through this model, researchers identify a form of social media hegemony.
Only one viewpoint is effectively communicated to a large audience, although multiple viewpoints exist on a controversial topic.
Significantly, recommender system algorithms accelerate the hegemony while exacerbating the misrepresentation of a minority viewpoint.

Some researchers confirm the ``filter bubble" in real social media and search engines.
Hussein et al.~\cite{hussein2020measuring} reveal the ``filter bubble" in the Top 5 and Up-Next recommendations topics (9/11 conspiracy theories, chemtrail conspiracy theory, flat earth, moon landing conspiracy theories, and vaccine controversies), except vaccine controversies on YouTube based on more than 56K videos.
Santos et al.~\cite{c2020there} also confirm the ``filter bubble" hypothesis on Twitter.
Although Twitter has tried to mitigate the ``filter bubble" by giving users more control, the actual performance of its personalization algorithms is still obscure.
Following an account out of sympathy for political topics can still cause Twitter Search to provide different results for polarized users.
Robertson et al.~\cite{robertson2018auditing} propose a Chrome extension to conduct a targeted algorithm audit for Google search.
The extension is designed to survey participants and collect search engine results.
They discover bias in search rankings on the desktop version of Google Search, where users' voting opinions are infected in policy competition.

The unexpected recommendation is also a useful method to solve the ``filter bubble".
It deviate significantly from users' previous expectations, thus surprising users by presenting them with new and previously unexplored items~\cite{li2020purs}.
Li and Tuzhilin~\cite{li2020latent} propose a new latent closure method to construct mixed utility functions based on differential geometry.
They use three common geometric structures to measure the closure of latent embeddings, including Hypersphere, Hypercube, and Convex Hull.
In contrast to previous models, they model user unexpectedness in the latent space and product embeddings, which can capture the complex relationships hidden between new recommendations and historical purchases.
Extensive experiments on three real-world datasets (Yelp, TripAdvisor, and Video) show that the model enables a significant increase in unexpected recommendations.
Li et al.~\cite{li2020purs} develop a personalized unexpected recommender system model.
The model implements an unexpected personalized function by providing multi-cluster modelling of user interests in the potential space through a self-focus mechanism and selecting appropriate unexpected recommendations.
These models enhance the novelty of recommendations while maintaining their accuracy, meeting the user's need for flexibility and individuality in the recommendation algorithm.

In conclusion, researchers directed their efforts towards identifying and revealing the presence of ``echo chambers" and ``filter bubbles". Whether these constrained recommendation methods are embraced or disregarded relies on individual preferences.
Employing unexpected recommendations is a common strategy to mitigate fairness issues, offering users diverse options and countering biased suggestions.

\subsection{Unfair Ranking}
Information access systems, including information retrieval and recommender systems, commonly rely on ranking to present outcomes pertinent to the user's information requirements~\cite{raj2022measuring}.
Ranking algorithms mistreat specific groups unfairly~\cite{fabris2023pairwise}.
Many online platforms usually adopt user-centric optimization where the ranking lists are generated by their estimated relevance~\cite{yang2023marginal}.
However, in these two-sided markets, the user-centric optimization ignores the profit for the item providers and causes fairness issues to item providers~\cite{kletti2022pareto}.
The extent to which users and item providers are receptive to the influence of rankings on their recommendations is contingent upon their interests and behavioural patterns~\cite{saito2022fair}.
The examples of unfair ranking are shown in Table~\ref{tab:RankingAlgorithms}.

\begin{table*}[]
  \small
	\centering
	\caption{The examples of the algorithmic fairness issues in ranking algorithms.}
	\label{tab:RankingAlgorithms}
	\begin{tabular}{p{2.5cm}| p{3.5cm} | p{6cm}}
		\hline
		\textbf{Algorithms} &
		\textbf{Types of bias} &
		\textbf{Examples} \\ \hline
		\textbf{Recommender Systems} &
		Position bias of offline evaluation &
		Users like to interact with the \textbf{top-positioned} items~\cite{saito2020unbiased} \\ \cline{2-3}
		\textbf{} &
		Position bias of online learning &
		Frequent update causes the \textbf{algorithms bias} over the Web~\cite{jadidinejad2020using}. \\ \cline{2-3}
		\textbf{} &
		Previous model bias of offline evaluation &
		The feedback dataset collection bias \textbf{towards the specific items} is supported by the deployment model~\cite{goren2019ranking}. \\ \cline{2-3}
		\textbf{} &
		Previous model bias of online learning &
		Feedback from the \textbf{past influences future} ranks in a specific form of online learning section information feedback~\cite{khenissi2020theoretical}. \\ \hline
		\textbf{Information Retrieval} &
		Position bias of offline evaluation &
		\textbf{Few-get-richer} effect is emerged when people have heterogeneous preferences for the classes of items~\cite{germano2019few}. \\ \cline{2-3}
		&
		Position bias of online learning &
		The \textbf{expected exposure} is biased of online information retrieval~\cite{diaz2020evaluating}. \\ \hline
	\end{tabular}
\end{table*}

\subsubsection{Recommender Systems}

Ideally, users' content choices are driven by their intentions, desires, and plans~\cite{wang2020collaborative}.
However, recommender systems may modify choices, promoting popular items and reinforcing users' historical behavior~\cite{baeza2020bias}.
In the learning-to-rank (LTR) algorithm,recommender systems adapt the ranking of items based on users' feedback~\cite{qin2020attribute}.
Unlike traditional recommender systems based on explicit feedback, modern systems use implicit feedback to train LTR models, such as dwell time, clicks, and purchases.
However, implicit feedback often faces bias, including previous model bias and position bias.
The previous model bias means that currently deployed models remarkably influence the training samples operated for future models~\cite{wang2021denoising}.
The position bias is another disadvantage of ranking algorithms, where users only attend to the items at higher ranking positions~\cite{chen2020efficient}.
Position bias exisits not only in item rankings but also in user rankings.
The trend for recommender systems is to provide better recommendations to mainstream users rather than non-mainstream users~\cite{li2021leave}.

The statistical parity fairness and the disparate impact fairness are two metrics of position fairness.
It is statistical parity fairness if documents from different groups have equal exposure.
It is the disparate impact fairness if documents from different groups have the same exposure proportion as the overall exposure proportion.
Zhu et al.~\cite{zhu2020measuring} formalize the concepts of these.
The first measure is the ranking-based statistical parity (RSP).
They assume a item attribute set $\mathcal{A} = \{a_1, a_2, ..., a_{|\mathcal{A}|}\}$ and define a function $G_{a_j}(i_1)$ to identify whether item $i_1$ belongs to $a_j$.
They calculate the probability distributions of different item groups as follows:
\begin{equation} \label{equ:ProbabilityDistribution}
\text{P}\left(\text{R@k} \mid a=a_{j}\right)=\frac{\sum_{u=1}^{|\mathcal{U}|} \sum_{i=1}^{k} G_{a_j}\left(r_{u, i}\right)}{\sum_{u=1}^{|\mathcal{U}|} \sum_{i \in \mathcal{I} \backslash \mathcal{H}_{i}} G_{a_j}(i)},
\end{equation}
where $\text{R@k}$ represents `being ranked in top-k', $\text{P}\left(\text{R@k} \mid a=a_j\right)$ is the probability of items in $a_j$ ranks in top-k. $\sum_{i=1}^{k} G_{a_j}\left(r_{u,i}\right)$ calculates the number of un-interacted items from $a_j$ rank in top-k for user $u$. $\sum_{i \in \mathcal{I} \backslash \mathcal{H}_{i}} G_{a_j}(i)$ calculates the number of un-interacted items belonging to $a_j$ for $u$.
Then they define the $\mathbf{RSP}@k$ as:
\begin{equation} \label{equ:RSP}
\mathbf{RSP}@k=\frac{\text{std}\left(\prod_{j=1}^{|\mathcal{A}|} \text{P}\left(\text{R@k} \mid a=a_{j}\right)\right)}{\text{mean}\left(\prod_{j=1}^{|\mathcal{A}|} \text{P}\left(\text{R@k} \mid a=a_{j}\right)\right)},
\end{equation}
where $\text{std}(\cdot)$ calculates the standard deviation, and $\text{mean}(\cdot)$ calculates the mean value.
The second measure is the ranking-based equal opportunity (REO), which combines RSP with equal odds.
They discover that the influential Bayesian personalized ranking model is susceptible to position bias and produces bias recommendations.
They then propose a new debiasing personalized ranking model incorporating adversarial learning.

The exposure and pairwise ranking accuracy are two metrics for measuring position bias.
Wang et al.~\cite{wang2021combating} discover the exposure bias in movie recommendations, where high ratings account for more observed ratings and under-represented low ratings.
Specifically, the percentage of high scores observed is $100\%$, higher than the percentage of low scores ($17\%$, nearly one-sixth).
Typically, systems recommend items to users sorted in descending order of predicted ratings.
Users' favor of interacting with popular items can mislead recommender systems tending to recommend popular items over tail items~\cite{saito2020unbiased}.
This biased feedback data leads the recommender systems to sub-optimal recommendations.
Wick et al.~\cite{wick2019unlocking} proposed semi-supervision to improve the accuracy and fairness of exposure while imparting beneficial properties of the unlabeled data to the classifier.
The model examines whether exposure is equality and considers whether current solutions may lead to under-compensation or over-compensation of equality approaches.
Singh and Joachims~\cite{singh2019policy} devise an LTR algorithm named Fair-PG-Rank (policy-gradient) for decreasing exposure bias.
Fair-PG-Rank identifies and neutralizes biased features to learn ranking functions efficiently under position fairness constraints.

The pairwise ranking accuracy is also essential for decreasing position bias, because in most debiasing methods, researchers only treat clicked documents as relevant.
These methods ignore the implicit bias created by treating non-clicked documents as irrelevant.
Wang et al.~\cite{wang2021non} propose a well-adjusted novel weighting scheme, named Propensity Ratio Scoring, which provides treatments on both users' clicks or non-clicks.
They apply this method to different ranking algorithms and show its wide applicability.
The scheme follows the conventional steps of synthesizing clicks on three benchmark LTR datasets (Yahoo, WEB10K, and MQ2007) to study PRS behaviour from different perspectives.
To verify the effectiveness of the method in an industrial setting, they also conduct experiments on large-scale data from Gmail searches.
This method decreases the implicit bias in non-clicked documents.
Click-through rate (CTR) is an attribute for training recommender algorithms, and some researchers focus on it to mitigate position bias.
Researchers have proposed position-independent and position-aware systems to estimate CTR.
The former systems assume that position $k$ does not affect the estimation of CTR.
Many recommender systems are position-aware.
Since ADs placed at different locations have different click probabilities, the location information needs to be considered during training.

To maximize revenue, retrieval bias and placement bias coexist in most real-world systems, including position-independent and position-aware systems.
Yuan et al.~\cite{yuan2020unbiased} propose a new counterfactual learning framework, where learning and evaluation are unbiased for positional and non-positional approaches.
The experiments in the ``Outbrain Click Prediction" competition and ``KKBox's Music Recommendation Challenge" competition confirm the framework's validity.
Zhuang et al.~\cite{zhuang2021cross} propose a cross-position attention mechanism to simulate the user's checking bias in clicks.
This attention mechanism discards the effect of product ranking, which removes position bias.
When users examine an item and find it relevant, they will click on it.
Their experiments demonstrate that their model could better mitigate position bias in click logs and learn more effective correlation scorers on multiple composites.

Significantly, some researchers discover a specific position bias, context bias, in click feedback.
Context bias means users prefer to click on items surrounded by non-similar items rather than those surrounded by similar items.
Wu et al.~\cite{wu2021unbiased} propose an Unbiased Learning to Rank with Combinational Propensity (ULTR-CP) framework to estimate context bias.
They instantiate the framework on e-commerce datasets to demonstrate that the model sufficiently decreases the context bias.

Another reason for unfair ranking is previous model bias.
The previous model bias creates closed-loop feedback in recommender systems, reinforcing the users' historical behaviour.
Jadidinejad et al.~\cite{jadidinejad2020using} study the influence of closed-loop feedback during the training process of recommendation systems. They use open-loop datasets, where randomly selected items are presented to users for feedback.
The solution is similar to ``unexpected recommendations".
The confounding effect of the previous model will be reduced if the deployment model explores a broader range of projects rather than developing highly related projects.
McInerney et al.~\cite{mcinerney2020counterfactual} propose a new counterfactual estimation method, dubbed as Reward Interaction Inverse Propensity Scoring.
It allows serial interactions in rewards with low variance asymptotically unbiasedly.
This method estimates sequences from causal graph specifications between interactions and rewards.
It avoids the overhead of full-parameter estimation associated with previous models and asymptotically decreases previous model bias.
Online learning with partial information feedback mainly suffers from solid influences of previous model bias.
The closed-loop feedback in recommender systems increases previous model bias in the continuous iterative systems process.
Khenissi et al.~\cite{khenissi2020theoretical} demonstrate that the iterative process of recommender systems increases the convergence properties of the user discovery because of previous model bias, based on the Movielens 1M dataset with $6040$ users, $3952$ items, and $1$ million ratings.

\subsubsection{Information Retrieval}

LTR algorithms are the core part of IR, for which many provably efficient methods have been recently proposed to analyze algorithmic bias in IR.
The position bias is the main cause of unfair ranking in IR.
Significantly, the position bias in IR leads to a ``few-get-richer" effect, which means that a few entries in a given class attract more traffic than other entries in the same class combined.
Germano et al.~\cite{germano2019few} present that the ``few-get-richer" effect occurs when people only click on top-ranked items and have heterogeneous preferences for item categories.
Intuition suggests that top-ranked items should be more relevant and trustworthy than other ones.
However, extensive researchers have confirmed that the ranking is often not informative about quality, especially in settings characterized by ``rich-get-richer" dynamics.
Popularity-based rankings create ``noise" in the ranking and systematic ranking bias.
When two distinct classes of items are trained in ranking algorithms, items from the smaller class have a better ranking than similar items from the larger class.
Researchers suggest adding clicks to the search engine algorithm, which is the best way to track the number of entries in each category~\cite{zhu2021popularity}.
Ideally, the ranking algorithm should use the popularity of different items neutrally.
Furthermore, some researchers advocate the principle of equal expected exposure.
When given a fixed information need, each item should receive equal exposure with another item of the same relevance grade.
Diaz et al.~\cite{diaz2020evaluating} argue that this principle is desirable for topical diversity and fair ranking.

Most of the research mitigates position bias in IR based on SVM-style methods.
The recently developed SVM-PropRank methods have shown that counterfactual inference techniques provably decrease the distorting effect of position bias in IR.
Based on these, Agarwal et al.~\cite{agarwal2019a} propose a framework for counterfactual LTR algorithms, which enables unbiased training for a broad class of additive ranking metrics (e.g., Discounted Cumulative Gain (DCG)) and a broad class of models (e.g., deep networks).
The framework includes two learning methods: SVM PropDCG and Deep PropDCG.
Based on experiments, they have found that SVM PropDCG outperform other SVM models, robustly decreasing propensity error descriptions.
The neural network in Deep PropDCG provides further improvements of DCG by making it more robust and efficient.
Gao and Shah~\cite{gao2020toward} focus on statistical parity fairness and disparate impact fairness.
They use 100 queries and the top 100 results per query from Google search data to reveal the existence of bias in top search results.
An entropy-based metric is proposed to measure the degree of bias.
They confirm that statistical parity fairness is highly correlated with diversity, while disparate impact fairness is not.

In summary, many researchers address the unfair ranking as a data sparsity issue.
These unfair ranking items disrupt the individual fairness of consumers and reduce the sustainability of providers. Enhancing the viability of providers also aids in diversifying recommendations.
However, similar to unfair ranking issues, concerns arise regarding the universal application of debiasing models to all users, as individuals' tolerance for long-tail problems varies.
The classification of users, particularly in the context of resolving these issues, remains an ongoing area of research.

\subsection{Long-tail Items}
\begin{figure}
	\centering
	\includegraphics[width=0.5\textwidth]{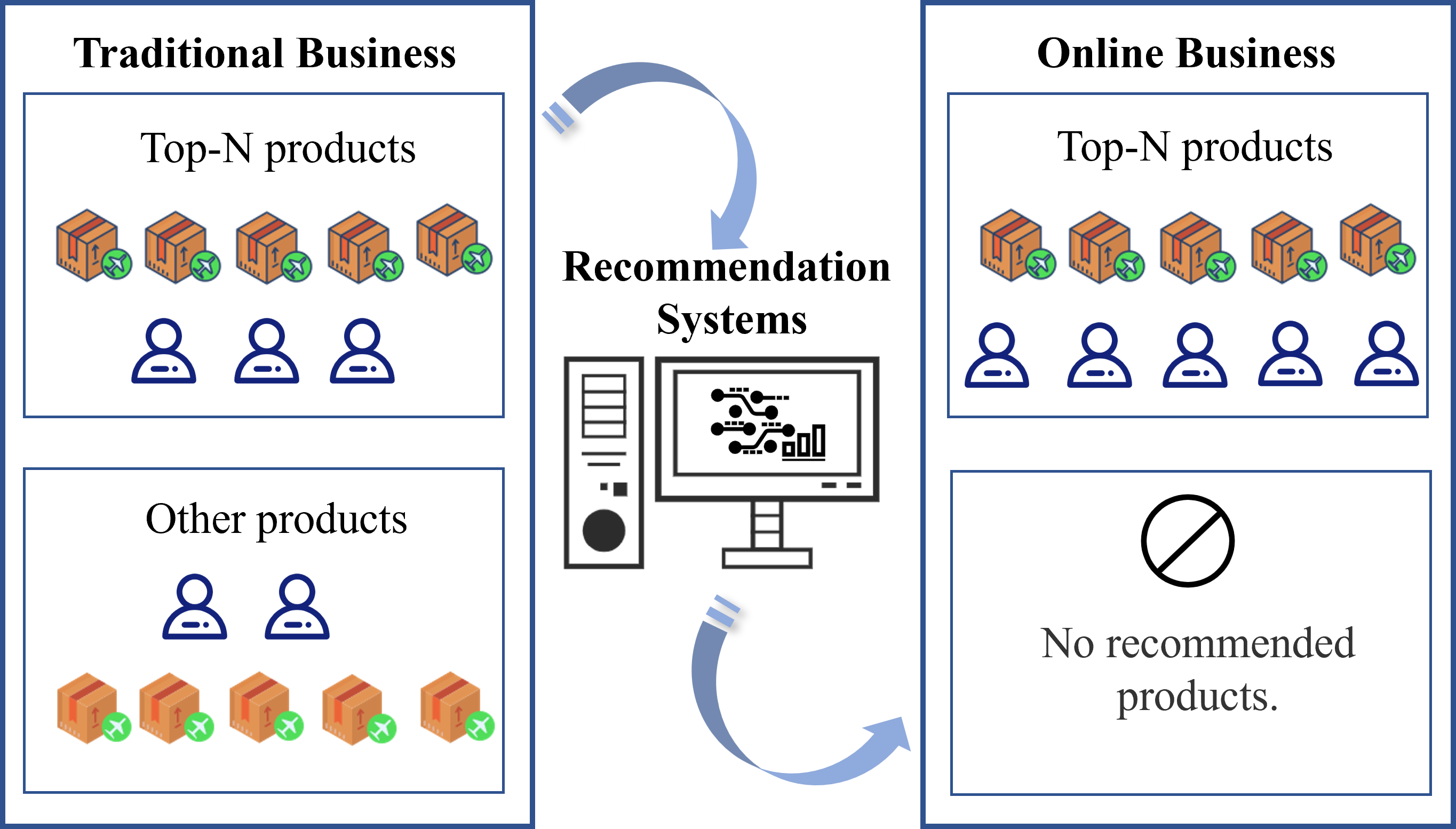}
	\caption{Long-tail items.}
	\label{fig:LI}
\end{figure}

In contrast to the relatively small number of top-ranked projects, many long-tail projects have fewer observations.
When using the Pareto principle (i.e., the $80/20$ $rule$), Zolaktaf et al.~\cite{zolaktaf2018generic} claim that long-tail items could be defined as items that generate the lower $20\%$ of observations.
Promotion of long-tail items increases overall coverage of the item space, increasing the fairness of ranking algorithms.
A lack of adequate feedback data to train ranking algorithms for long-tail items leads to long-tail bias.
The top-ranked shops could limit the search results and degrade the users' experience.
The schematic is shown in Fig~\ref{fig:LI}.

Some researchers study the long-tail item's problem as a data sparsity problem because long-tail bias lacks feedback data.
The graph neural networks (GNNs) enhance the node representations, which can be used as a valuable method for solving the data sparsity problem~\cite{xia2021graph}.
Niu et al.~\cite{niu2020a} propose a dual heterogeneous graph attention network (DHGAT) based on GNNs, which gives the long-tail items a fair chance to search and discover.
DHGAT is a deep end-to-end framework with a two-tower architecture based on graphically enhanced ID representations and textual representations.
It better describes user intent and retrieves personalized store search results.
After conducting a large-scale offline evaluation and online A/B testing, the results show that DHGAT has significant advantages over solid baselines and effectively leverages heterogeneous graphs.
Chen et al.~\cite{chen2020esam} propose an entire space adaptation model (ESAM) to address the data sparsity problem of non-displayed items (most are long-tail items) from the perspective of domain adaptation (DA).
Specifically, they design the attribute correlation alignment by considering the correlation between high-level attributes of items to achieve distribution alignment.
ESAM feeds long-tail items into the stream with three constraints to improve the DA process, including domain adaptation with attribute correlation alignment, central-wise clustering for source clustering, and self-training for target clustering.
Offline experiments on two public datasets and a Taobao industrial dataset demonstrate that ESAM achieves state-of-the-art performance.

Collaborative filtering (CF) methods have been widely used in recommender systems.
Most existing hybrid CF methods mitigate the data sparsity problem through integrating side information.
However, the process of integrating side information is computationally expensive.
Wang et al.~\cite{wang2019enhancing} develope a cost-effective CF model named AugCF, based on Conditional Generative Adversarial Nets.
The model is established on data augmentation and neural networks. A novel discriminator loss function is proposed to handle different tasks in different training phases to enjoy the benefits of end-to-end training.
The model optimises the generator using Gumbel-Softmax to overcome the gradient block problem while reducing the variance, which decreases the cost of integrating side information without decreasing the performance of debiasing.

The cold-start problem is defined as recommendation without any available historical feedback in the target system. Some researchers dedicated to simultaneously alleviating the long-tail items and cold-start. For example, Li et al.~\cite{li2019both} propose novel approach based on an iterative optimization algorithm.
They have found it beneficial to consider long-tail items in the cold-start process.
In the proposed method, items of interests are divided into two parts: a low-rank part for short-head items and a sparse part for long-tail items.
Then, they handle short-headed items and long-tail items independently in traditional recommender systems, which solves the long-tail bias and cold-start problem.
Side information warms up recommender systems when history feedback is not available in recommender systems.
Liu et al.~\cite{liu2021social} propose a novel joint recommendation model (InSRMF), based on the Indirect Social Relations detection and Matrix Factorization collaborative filtering.
In the model, the latent user factors simultaneously and seamlessly mine the user's personal preferences and social group characteristics.
InSRMF identifies the potential feature space while capturing user overlap, community structure and individual preferences for items.
InSRMF integrates indirect social relationship identification, potential users, and project factor identification, which could recommend all users, including near-cold-start users, pure-cold-start users, long-tail items.
Yin et al.~\cite{yin2020learning} use learning transferrable parameters from both optimization and feature perspectives to solve long-tail and cold-start problems.
Specifically, they propose a gradient alignment optimizer that encourages knowledge transfer from an optimization perspective.
They introduce an adversarial training method to learn frequency-agnostic sequence embeddings while facilitating knowledge transfer from a feature perspective.

Some researchers are also working on mitigating provider long-tail bias.
Mladenov et al.~\cite{mladenov2020optimizing} optimize the social welfare of long-tail shops to increase the viability of providers through mitigating the matching bias.
Their linear programming-based approach improves providers' and users' welfare over short-sighted strategies.

In summary, most researchers solve the long-tail bias as a data sparsity problem.
Some researchers mitigate long-tail bias while resolving the cold-start problem.
These long-tail items break the individual fairness of consumers and decrease the viability of providers.
Increasing the viability of providers also facilitates the improvement of recommendation diversity.
Similar to the issue of unfair ranking, there are concerns about the applicability of debiasing models to all users, as well as individual differences in tolerance for long-tail problems.
User classification, especially in the context of addressing these problems, is an area of ongoing research.

\section{Open Challenges and Implications} \label{sec:OCaI}

\subsection{Beyond Binary Sensitive Attributes}
Current fairness-aware algorithms primarily focus on addressing biases related to binary sensitive attributes~\cite{maheshwari2023fairgrad}.
For example, racial attributes include Asians, African-Americans, Whites, and others, which are often treated separately in current research, with limited consideration given to their combined effects.
While some recent studies have begun to address bias with multiple sensitive attributes, it is worth noting that the performance of fairness-aware methods for binary sensitive attributes surpasses that of their counterparts designed for multiple attributes~\cite{d2023debiaser}.
A promising future direction in this field involves harmonizing fairness considerations for multiple sensitive attributes while maintaining satisfactory performance levels.

\subsection{Debiasing in Privacy Data}
As individuals grow more cognizant of privacy concerns, they exhibit hesitancy in sharing sensitive information about their data~\cite{dai2021say}.
A significant portion of debiasing research regarding fairness relies on complete datasets~\cite{lahoti2020fairness}.
However, in real-world scenarios, the availability of sensitive data is often limited or obscured due to privacy safeguards.
Consequently, debiasing with constrained data poses a formidable challenge when operating with constrained or incomplete data. While approaches like joint learning and differential privacy have shown promise in preserving data privacy, their suitability for addressing biases and achieving fairness in algorithms must be reevaluated.
Ensuring legal tolerant fairness while upholding the protection of sensitive attributes emerges as a novel domain at the intersection of two ethical dimensions of machine learning~\cite{dai2021say}.

\subsection{Defining Ethical Regulation in Different Fields}
Ensuring ethical tolerance and fairness necessitates the identification and consideration of specific ethical regulations that pertain to each case. The definition of these attributes varies across different fields of expertise, highlighting the importance of context in assessing fairness.
Algorithmic fairness is not solely a concern within computer science; it extends to other disciplines such as social science, medicine, and economics. Recognizing this interdisciplinary nature is crucial for a comprehensive understanding and effective implementation of fairness principles. For instance, in economics, ethical regulations protect users' location information, which researchers use to make critical financial decisions like loan approvals, irrespective of the users' living conditions. Understanding such ethical regulations and their implications is paramount in ensuring fairness in decision-making processes~\cite{rueden2023informed}. This highlights the need for interdisciplinary research that encompasses both domain-specific knowledge and ethical considerations.

\subsection{Debiasing for Personalized Requirements}
Personalized tolerance fairness prioritizes meeting the individual needs of users. Some users may prefer recommendations based on statistically popular items, while others seek diversity and constant surprise in their recommendations. In such cases, algorithms should be personalized to align with the specific preferences of users.
In real-world environments, personalized attributes are derived from user preferences, which may introduce biases~\cite{bose2019compositional}.
For example, , female users might prioritize fairness more than male users, potentially resulting in different personalized tolerance fairness requirements.
A research direction in personalized tolerance fairness involves striving to meet the algorithmic expectations of individual users to the greatest extent possible while considering the associated costs and feasibility.

\section{Conclusion} \label{sec:Con}
In this survey, we provide an overview of algorithmic fairness, emphasizing the concept of tolerance. We explore various scenarios where fairness issues in algorithms could lead to potential harm and review several techniques that promote fairness in algorithmic decisions. Our analysis particularly examines discrimination within fairness-aware algorithms and underscores the non-negotiable stance on legal tolerance due to the significant impact of social discrimination on individuals' lives. Ethical tolerance is considered in the context of policy competitions and non-competitive pricing, despite the absence of specific legal frameworks. However, prevailing social norms strongly oppose prejudice, often leading to widespread condemnation.

We highlight the urgent need for policies or methodologies that can effectively mitigate the adverse consequences of such prejudices. Personal tolerance is discussed in terms of restricted access to information, unfair rankings, and the handling of long-tail items, advocating for algorithms that protect user content choices based on their intentions, desires, and plans.

This article identifies four promising future research directions in algorithmic fairness: exploring beyond binary sensitive attributes, debiasing private data, defining ethical regulations across various fields, and debiasing for personalized requirements. Advancing research in these areas aims to enhance fairness and reduce the negative impacts of algorithmic decision-making.

\section*{Acknowledgment}
The authors would like to thank Vidya Saikrishna (Federation University Australia), Yanjie Fu (Arizona State University), and Jiliang Tang (Michigan State University) for their valuable comments and suggestions. This work was partially supported by National Natural Science Foundation of China under Grant No. 72204037 and Dalian Science and Technology Talent Innovation Support Program (Youth Science and Technology Star) (Project No. 2022RQ055)

\bibliographystyle{ACM-Reference-Format}
\bibliography{Algorithmic_Fairness}

\end{document}